\documentclass[openacc]{article}
\usepackage[a4paper, margin=1in]{geometry}

\usepackage[most]{tcolorbox}
\usepackage{xcolor}
\usepackage{nicefrac}
\usepackage[normalem]{ulem}
\usepackage{physics}
\usepackage{amsfonts}
\usepackage{amssymb}
\usepackage{stmaryrd}
\usepackage{mathrsfs}
\usepackage{bm}
\usepackage{soul}
\usepackage{titlesec} 
\usepackage{url}
\usepackage{subcaption}
\usepackage{hyperref}

\mathcode`@="8000 
{\catcode`\@=\active\gdef@{\mkern1mu}}

\newcommand*\oline[1]{%
  \vbox{%
    \hrule height 0.5pt%                  
    \kern0.25ex%                          
    \hbox{%
      \kern-0.1em%                        
      \ifmmode#1\else\ensuremath{#1}\fi%  
      \kern-0.1em%                        
    }
  }
}

\newcommand{\diff}[2]{\frac{\mathrm{d} #1}{\mathrm{d} #2}}

\newcommand{\transpose}{\mathrm{T}}

\sethlcolor{green!20}

\begin{document}

%%%% Article title to be placed here
\title{The Role of Compressibility in Modified Quasi-Linear Viscoelasticity: A Comparison of Simple Shear and Torsion}
\date{}

\author{Valentina Balbi$^1$ and Griffen Small$^2$}

\footnotetext[1]{Valentina Balbi: School of Mathematical and Statistical Sciences, University of Galway, College Road, Galway, H91 TK33, Ireland. Email: vbalbi@universityofgalway.ie}
\footnotetext[2]{Griffen Small: Department of Mechanical and Manufacturing Engineering, University of Calgary, 2500 University Drive NW, Calgary, Alberta, T2N 1N4, Canada}

\maketitle
\begin{abstract}
We investigate the role of compressibility in the modified quasi-linear viscoelastic (MQLV) constitutive framework for soft solids at finite strain, where shear and bulk responses are governed by distinct relaxation functions. Analytical and semi-analytical results are derived for simple shear and torsion, under incompressible and slightly compressible assumptions.
We show that compressibility affects the response only when volume changes occur: under isochoric deformations, the bulk contribution vanishes, while even small deviations from isochoricity significantly alter the normal response. Shear stress and torque are largely insensitive to compressibility, whereas normal stress and axial force exhibit pronounced sensitivity due to the coupling between shear and bulk relaxation.
We further demonstrate that volumetric effects interact with the Poynting effect: in simple shear they oppose each other, reducing relaxation, while in torsion they reinforce each other, enhancing it. These trends agree with brain tissue experiments but reveal limitations of the slightly compressible model for highly compressible materials, such as agarose gels.
Overall, the results emphasise the importance of accounting for compressibility in modelling normal stress responses and motivate the development of fully compressible formulations and numerical implementations.

\end{abstract}

\section{Introduction}
\label{sec: introduction}

The mechanical behaviour of soft solids, including biological tissues and polymeric gels, is characterised by large deformations, strong non-linearity and pronounced time-dependent effects. Capturing these features within a unified constitutive framework remains a central problem in continuum mechanics, with important implications for biomechanics and engineering applications. A wide range of constitutive approaches to non-linear viscoelasticity have been proposed, broadly classified into integral-type, differential and internal-variable models. Integral formulations, based on the theory of materials with memory, express the stress as a functional of the entire deformation history, as in the classical Pipkin--Rogers and K--BKZ models~\cite{wineman09}. By contrast, differential models relate stress to strain and its time derivatives, but often struggle to capture long-term relaxation behaviour~\cite{berjamin21}. Internal-variable approaches, grounded in thermodynamics, introduce additional state variables to describe dissipative mechanisms and are closely related to multiplicative frameworks, in which the deformation gradient is decomposed into elastic and viscous parts~\cite{reese96,simo87,sidoroff74}.

Within this broad framework, quasi-linear viscoelasticity (QLV), originally introduced by Fung~\cite{fung93}, has become one of the most widely used models in biomechanics due to its simplicity and ability to incorporate non-linear elasticity and time dependence. QLV is an integral-type model, in which the stress is expressed as the convolution of a scalar relaxation function with an instantaneous elastic response~\cite{fung93}. While this separable structure is computationally attractive and has been successfully applied to a wide variety of soft tissues, it also imposes significant limitations. In particular, the use of a single scalar relaxation function restricts the model to incompressible behaviour and prevents a consistent description of general three-dimensional viscoelastic responses~\cite{depascalis14}.

To overcome these limitations, the modified quasi-linear viscoelastic (MQLV) framework was developed by De Pascalis \textit{et al.}~\cite{depascalis14}, in which the relaxation behaviour is represented by a tensorial relaxation function. This formulation enables a consistent decomposition of the stress into volumetric and deviatoric contributions, each governed by its own relaxation function, thereby ensuring that the model reduces to classical linear viscoelasticity in the small-strain limit. The MQLV model has been applied to study various deformation modes, including uniaxial tension~\cite{depascalis14}, simple shear~\cite{depascalis15}, torsion~\cite{small25,righi21} and inflation~\cite{depascalis18}, and has also been extended to anisotropic materials~\cite{balbi23,balbi18}. In particular, the MQLV approach provides a natural way to incorporate physically distinct relaxation mechanisms associated with the bulk and shear responses of a material.

Despite these advances, most applications of MQLV theory have focused on incompressible materials, reflecting the common assumption that soft solids undergo negligible volume changes. However, both experimental evidence and theoretical considerations indicate that all real materials are compressible to some extent. For example, skeletal muscle exhibits measurable deviations from isochoric behaviour due to fluid exchange and vascular effects during deformation~\cite{causey12}; ligaments and tendons undergo significant volume changes under tensile loading, as evidenced by strain-dependent Poisson's ratios~\cite{swedberg14} and brain tissue displays coupled volumetric and mechanical responses arising from its highly hydrated multiphasic structure~\cite{greiner24}.

Motivated by these observations, compressible viscoelastic behaviour has been investigated within several constitutive frameworks, including internal-variable formulations at finite strain~\cite{holzapfelGasser02,holzapfelSimo96}, phenomenological models for polymers~\cite{cheng10} and hereditary formulations used in computational mechanics~\cite{abaqus}. While these models incorporate compressibility through the volumetric response, the time-dependent behaviour is often described by a single effective relaxation function or by relaxation processes that act uniformly across the stress. For instance, internal-variable models based on short- and long-term mechanisms~\cite{ahsanizadeh15} introduce multiple time scales, but these are embedded within differential evolution equations and do not correspond to independent tensorial relaxation functions associated with distinct physical modes of deformation. Moreover, poroviscoelastic models attribute volume changes to fluid transport rather than intrinsic solid relaxation~\cite{greiner24}, leaving open the question of how multiple relaxation mechanisms interact in compressible viscoelastic solids.

\begin{figure}[t!]
    \centering
    \begin{subfigure}{0.49\textwidth}
        \centering
        \includegraphics[width=\linewidth]{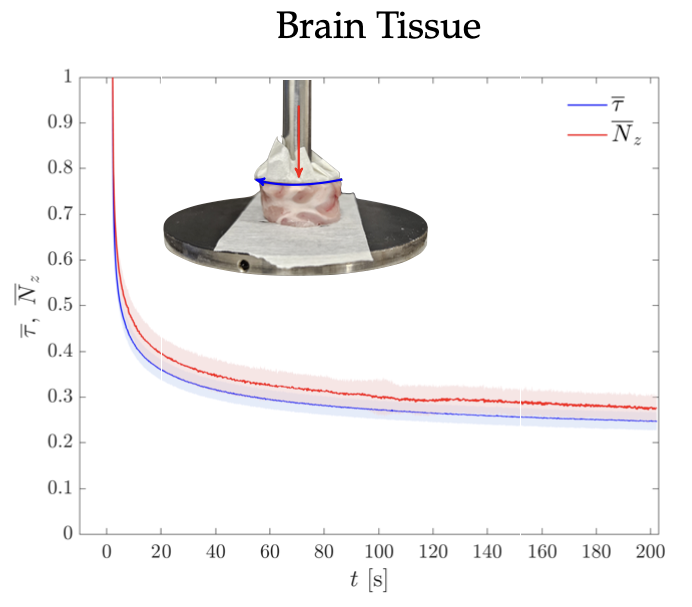}
        \caption{}
        \label{fig: brain data}
    \end{subfigure}
    \hfill
    \begin{subfigure}{0.49\textwidth}
        \centering
        \includegraphics[width=\linewidth]{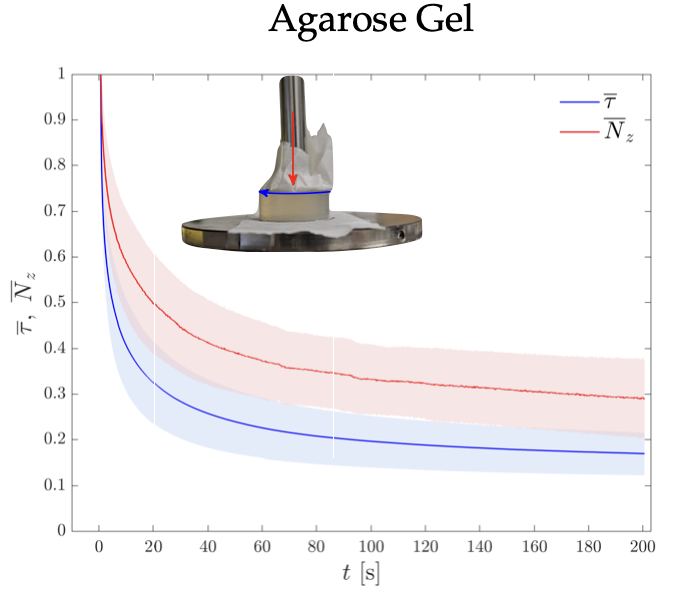}
        \caption{}
        \label{fig: gel data}
    \end{subfigure}
    \caption{Normalised torque $\overline{\tau}$ and axial force $\overline{N}_z$ from torsion tests performed on (a)~10 brain tissue samples and (b)~10 agarose gel samples of radius $12.5\,\mathrm{mm}$ and height $\sim 10\,\mathrm{mm}$. Data are presented as mean (solid curves) and standard deviation (surrounding colour bands). The twist rate $\dot{\phi}_0=40\,\mathrm{rad}\,\mathrm{m}^{-1}\,\mathrm{s}^{-1}$ was the same in both cases, but the final value of the twist differed, with $\phi_0=88\,\mathrm{rad}\,\mathrm{m}^{-1}$ for brain tissue and $\phi_0=25\,\mathrm{rad}\,\mathrm{m}^{-1}$ for agarose gel.}
    \label{fig:Normalised experimental data}
\end{figure}

In addition to these modelling considerations, experimental observations provide further evidence that multiple relaxation mechanisms may be active in soft solids. In particular, torsion experiments on ovine brain tissue~\cite{small25,BrainTorsionData} and $2\%$ w/v agarose gel~\cite{GelTorsionData} reveal that the normalised relaxation curves of torque and axial force do not coincide, as shown in Figure~\ref{fig:Normalised experimental data}. This discrepancy is especially pronounced in agarose gels and suggests a competition between distinct relaxation processes. Such behaviour cannot be captured by classical QLV models with a single relaxation function but is naturally accommodated within the MQLV framework through the presence of multiple relaxation mechanisms.

The aim of this paper is therefore to investigate the effect of compressibility on the modified quasi-linear viscoelastic model for soft solids. We consider two constitutive settings---slightly compressible and incompressible---and analyse their predictions for two canonical deformations: simple shear and torsion. For simple shear, we examine both isochoric and nearly-isochoric deformations, highlighting the influence of volumetric relaxation on both shear and normal stress components. For torsion, we derive analytical results in the incompressible case and develop a perturbative approach for slightly compressible solids, enabling a systematic assessment of the roles of bulk and shear relaxation in determining the torque and axial force response.

%%%%%%%%%%%%%%%%%%%%%%%%%%%%%%%%%%%%%%%%%%%%%%%%%%%%%%
\section{Modified quasi-linear viscoelasticity}
\label{sec: mqlv}

In this section, we outline the equations governing the deformation of an isotropic viscoelastic soft solid within the MQLV framework. We consider the slightly compressible and incompressible cases and, for both, formulate the corresponding model and specify the associated constitutive assumptions for the relaxation and strain energy functions. For a more detailed and comprehensive account of the isotropic MQLV theory, including derivations of the  associated equations, we refer the reader to~\cite{Small25PhD,balbi23,righi21,depascalis14}.

%%%%%%%%%%%%%%%%%%%%%%%%%%%%%%%%%%%%%%%%%%%%%%%%%%%%%%
\subsection{Compressible form}
\label{sec: compressible mqlv}

In the large-deformation regime, the motion of a solid continuum is described by a deformation $\bm{\chi}$ that transforms a material point with position vector $\bm{X}$ in the undeformed configuration at time $t=0$ to a point with position vector $\bm{x}(t)$ in the deformed configuration, such that $\bm{x}(t)=\bm{\chi}(\bm{X},t)$. The associated deformation gradient tensor is defined as $\bm{F}=\partial \bm{x} \slash \partial \bm{X}$, from which various kinematic tensors can be derived, including the right Cauchy--Green deformation tensor $\bm{C}=\bm{F}^{\transpose}\bm{F}$ and the left Cauchy--Green deformation tensor $\bm{B}=\bm{F}\bm{F}^{\transpose}$.

In the isotropic MQLV framework, the constitutive equation in its most general form can be expressed in terms of the second Piola--Kirchhoff stress tensor as follows~\cite{Small25PhD,balbi23}: 
\begin{equation}
    \bm{\Pi}(t)=J(t)\bm{F}(t)^{-1}\left(\mathbb{G}(0)\bm{:}\bm{T}^{\mathrm{e}}(t)\right)\bm{F}(t)^{-\transpose}+\int_{0}^{t}J(s)\bm{F}(s)^{-1}\left(\mathbb{G}^{\prime}(t-s)\bm{:}\bm{T}^{\mathrm{e}}(s)\right)\bm{F}(s)^{-\transpose}\dd{s}, 
    \label{MQLV constitutive equation for 2nd PK}
\end{equation} 
where $J=\mathrm{det}\,\bm{F}$ is the volume ratio, $\bm{T}^{\mathrm{e}}$ is the elastic Cauchy stress tensor and the prime denotes a derivative with respect to the argument of the function. The reduced relaxation tensor $\mathbb{G}=\sum_{i=1}^{2}G_i\mathbb{I}_i$ is a fourth-order tensor whose components $G_i$ are the relaxation functions that characterise the viscoelastic behaviour of the material. By choosing an appropriate tensor basis $\{\mathbb{I}_1,\mathbb{I}_2\}$ for $\mathbb{G}$, these functions can be identified with the different time-dependent mechanical moduli of the material. In particular, as shown in~\cite{Small25PhD,balbi23}, by expressing the reduced relaxation tensor in terms of the basis:
\begin{equation}
    \left(\mathbb{I}_1\right)_{ijkl}=\frac{1}{3}\delta_{ij}\delta_{kl} \qquad \text{and} \qquad \left(\mathbb{I}_2\right)_{ijkl}=\frac{1}{2}(\delta_{ik}\delta_{jl}+\delta_{il}\delta_{jk})-\frac{1}{3}\delta_{ij}\delta_{kl}, 
    \label{G basis tensors}
\end{equation}
with $\delta_{ij}$ denoting the Kronecker delta, we can identify the associated components as the normalised bulk and shear moduli:   
\begin{equation}
    G_1(t)=\frac{\kappa(t)}{\kappa_0} \qquad \text{and} \qquad G_2(t)=\frac{\mu(t)}{\mu_0},
    \label{Components of G}
\end{equation}
where $\kappa(t)$ and $\mu(t)$ denote the relaxation functions for the bulk and shear moduli, respectively, while $\kappa(0)=\kappa_0$ and $\mu(0)=\mu_0$ denote the corresponding instantaneous moduli. 
With this same choice of basis, the double contractions between $\mathbb{G}$ and $\bm{T}^{\mathrm{e}}$ in~\eqref{MQLV constitutive equation for 2nd PK} naturally split the elastic response into separate hydrostatic and deviatoric parts~\cite{Small25PhD,balbi23}. Altogether, this yields a particularly tractable form of the MQLV constitutive equation, expressed in terms of two distinct and physically meaningful relaxation functions for the bulk and shear moduli~\cite{Small25PhD,balbi23}: 
\begin{align}
    \bm{\Pi}(t)&=\left(\bm{\Pi}^{\mathrm{e}}_{\mathrm{H}}(t)+\frac{1}{\kappa_0}\int_{0}^{t}\kappa^{\prime}(t-s)\bm{\Pi}^{\mathrm{e}}_{\mathrm{H}}(s)\dd{s}\right) \nonumber \\
    &+\left(\bm{\Pi}^{\mathrm{e}}_{\mathrm{D}}(t)+\frac{1}{\mu_0}\int_{0}^{t}\mu^{\prime}(t-s)\bm{\Pi}^{\mathrm{e}}_{\mathrm{D}}(s)\dd{s}\right),
    \label{MQLV decomposition of Pi}
\end{align}
where we have defined the following terms:
\begin{equation}
\left\{
    \begin{alignedat}{3}  
    &\bm{\Pi}^{\mathrm{e}}_{\mathrm{H}}=J\bm{F}^{-1}\bm{T}^{\mathrm{e}}_{\mathrm{H}}\bm{F}^{-\transpose}, \qquad 
    &&\bm{\Pi}^{\mathrm{e}}_{\mathrm{D}}=J\bm{F}^{-1}\bm{T}^{\mathrm{e}}_{\mathrm{D}}\bm{F}^{-\transpose},  \\
    &\bm{T}^{\mathrm{e}}_{\mathrm{H}}=\frac{1}{3}\left(\mathrm{tr}\,\bm{T}^{\mathrm{e}}\right)\bm{I},  \qquad 
    &&\bm{T}^{\mathrm{e}}_{\mathrm{D}}=\bm{T}^{\mathrm{e}}-\frac{1}{3}\left(\mathrm{tr}\,\bm{T}^{\mathrm{e}}\right)\bm{I},
    \end{alignedat}
    \label{MQLV Piola transforms and hyd/dev splits}
\right.
\end{equation}
with $\bm{I}$ being the identity tensor. Here, $\bm{T}^{\mathrm{e}}_{\mathrm{H}}$ and $\bm{T}^{\mathrm{e}}_{\mathrm{D}}$ denote the hydrostatic and deviatoric parts, respectively, of the elastic Cauchy stress tensor $\bm{T}^{\mathrm{e}}$, while $\bm{\Pi}^{\mathrm{e}}_{\mathrm{H}}$ and $\bm{\Pi}^{\mathrm{e}}_{\mathrm{D}}$ denote the corresponding second Piola--Kirchhoff stress tensors. Applying the inverse Piola transformation $\bm{T}=J^{-1}\bm{F}\bm{\Pi}\bm{F}^{\transpose}$ to the second Piola--Kirchhoff stress tensor~\eqref{MQLV decomposition of Pi}, we obtain the  corresponding MQLV constitutive equation for the (viscoelastic) Cauchy stress tensor:
\begin{align}
    \bm{T}(t)&=J(t)^{-1}\bm{F}(t)\left(\bm{\Pi}^{\mathrm{e}}_{\mathrm{H}}(t)+\frac{1}{\kappa_0}\int_{0}^{t}\kappa^{\prime}(t-s)\bm{\Pi}^{\mathrm{e}}_{\mathrm{H}}(s)\dd{s}\right)\bm{F}(t)^{\transpose} \nonumber  \\   
    &+J(t)^{-1}\bm{F}(t)\left(\bm{\Pi}^{\mathrm{e}}_{\mathrm{D}}(t)+\frac{1}{\mu_0}\int_{0}^{t}\mu^{\prime}(t-s)\bm{\Pi}^{\mathrm{e}}_{\mathrm{D}}(s)\dd{s}\right)\bm{F}(t)^{\transpose}.
    \label{Cauchy viscoelastic stress}
\end{align}

For soft solids, the simplest choice for the time-dependent relaxation functions $\kappa(t)$ and $\mu(t)$ in~\eqref{Cauchy viscoelastic stress} are one-term Prony series of the form:
\begin{equation}
    \kappa(t)=\kappa_{\infty}+\left(\kappa_0-\kappa_{\infty}\right)\mathrm{e}^{-t \slash \tau_{\mathrm{H}}} \qquad \text{and} \qquad \mu(t)=\mu_{\infty}+\left(\mu_0-\mu_{\infty}\right)\mathrm{e}^{-t \slash \tau_{\mathrm{D}}},
    \label{Prony series}
\end{equation}
where $\kappa_{\infty}$ and $\mu_{\infty}$ denote the long-time bulk and shear moduli, respectively, while $\tau_{\mathrm{H}}$ and $\tau_{\mathrm{D}}$ denote the associated relaxation times for the bulk and shear response. In practice, however, soft solids often require multi-term Prony series to accurately fit stress relaxation data~\cite{matjeka26,small25,ed21,karimi16}. Nevertheless, one-term series are typically sufficient to capture the main qualitative features of the relaxation response~\cite{balbi23,balbi18,depascalis18,depascalis15,depascalis14} and result in a considerably simpler form of the MQLV constitutive equation than would be obtained using multi-term series.

In addition to the choice of relaxation functions, constitutive assumptions must also be specified for the instantaneous response. Non-linear elastic behaviour can be readily incorporated in~\eqref{Cauchy viscoelastic stress} through the elastic Cauchy stress tensor $\bm{T}^{\mathrm{e}}$ by adopting a hyperelastic constitutive framework.  In this setting, $\bm{T}^{\mathrm{e}}$ is expressed in terms of a strain energy function $W$, which, for isotropic compressible materials, depends only on the principal invariants $I_1$, $I_2$ and $I_3$ of $\bm{C}$, defined as:
\begin{equation}
    I_{1}=\mathrm{tr}\,\bm{C}, \qquad
    I_{2}=\frac{1}{2}\left(I^2_1-\mathrm{tr}\,\bm{C}^2\right) \quad \text{and} \quad  
    I_{3}=\mathrm{det}\,\bm{C}=J^2.
    \label{Principal invariants}
\end{equation}
This form of the strain energy function leads to the following representation formula for $\bm{T}^{\mathrm{e}}$~\cite{destrade25, anand20, atkin13, holzapfel00, ogden97}:
\begin{equation}
    \bm{T}^{\mathrm{e}}=\beta_{0}\bm{I}+\beta_{1}\bm{B}+\beta_{-1}\bm{B}^{-1},
    \label{Elastic compressible constitutive equation}
\end{equation}
where the response functions $\beta_0$, $\beta_1$ and $\beta_{-1}$ are given by:
\begin{equation}
\left\{
    \begin{aligned}
        &\beta_0=2J^{-1}\left(I_2W_2+I_3W_3\right), 
        \\
        &\beta_1=2J^{-1}W_1, 
        \\
        &\beta_{-1}=-2JW_2,
    \end{aligned}
    \label{Material response functions for compressible material}
\right.
\end{equation}
with $W_i=\partial W \slash \partial I_i$ for $i=1,2,3$. Taking the trace of~\eqref{Elastic compressible constitutive equation} and making use of the identity $\mathrm{tr}\,\bm{B}^{-1}=I_2 \slash I_3$ gives:
\begin{equation}
    \mathrm{tr}\,\bm{T}^{\mathrm{e}}=3\beta_0+I_1\beta_1+\frac{I_2\beta_{-1}}{I_3}.
    \label{Trace of Te}
\end{equation}
Using~\eqref{Elastic compressible constitutive equation} and~\eqref{Material response functions for compressible material}, together with~\eqref{Trace of Te}, the elastic second Piola--Kirchhoff stress tensors in~\eqref{MQLV Piola transforms and hyd/dev splits} take the form:  
\begin{equation}
	\left\{
	\begin{aligned}
	&\bm{\Pi}^{\mathrm{e}}_{\mathrm{H}}=2\left(\frac{2}{3}I_2W_2+\frac{1}{3}I_1W_1+I_3W_3\right)\bm{C}^{-1}, 
        \\	
	&\bm{\Pi}^{\mathrm{e}}_{\mathrm{D}}=2\left(W_1\bm{I}+\frac{1}{3}\left(I_2W_2-I_1W_1\right)\bm{C}^{-1}-I_3W_2\bm{C}^{-2}\right).
	\end{aligned}
	\label{Second Piola-Kirchhoff stress tensors compressible}
    \right.
\end{equation}
Substituting~\eqref{Second Piola-Kirchhoff stress tensors compressible} into~\eqref{Cauchy viscoelastic stress} then yields the MQLV constitutive equation for an isotropic compressible viscoelastic solid with instantaneous hyperelastic response. 

%%%%%%%%%%%%%%%%%%%%%%%%%%%%%%%%%%%%%%%%%%%%%%%%%%%%%%
\subsection{Slightly compressible form}
\label{sec: slightly compressible mqlv}

A compromise between perfect incompressibility and full compressibility is slight-compressibility. A material is classified as slightly compressible (or nearly incompressible) when its instantaneous bulk modulus is finite but several orders of magnitude larger than its instantaneous shear modulus~\cite{horgan09}. The most common approach to modelling slight compressibility in isotropic hyperelastic solids is to assume, largely for mathematical simplicity, that the strain energy function can be additively decomposed into uncoupled isochoric and volumetric contributions as follows~\cite{anand20,destrade11,horgan10,horgan07,holzapfel00}: 
\begin{equation}
    W=\psi(\bar{I}_1,\bar{I}_2)+U(J),
    \label{Slightly compressible strain energy}
\end{equation}
where $\bar{I}_1=\mathrm{tr}\,\overline{\bm{C}}$ and $\bar{I}_2=\mathrm{tr}\,\overline{\bm{C}}^{-1}$ are the first and second principal invariants of the modified right Cauchy--Green deformation tensor $\overline{\bm{C}}=J^{-2 \slash 3}\bm{C}$, respectively. The first term $\psi$ represents the isochoric contribution and is obtained from the corresponding incompressible strain energy function by replacing $I_1$ and $I_2$ with the modified invariants $\bar{I}_1$ and $\bar{I}_2$, respectively. The second term $U$, which depends only on $J$, measures volume change and thus accounts for deviations from perfect incompressibility. This implementation of slight compressibility is widely used in finite element packages, including FEBio~\cite{FEBioTheoryManual}, Abaqus~\cite{AbaqusTheoryManual} and Ansys~\cite{AnsysTheoryReference}.

While the slightly compressible form of the strain energy function~\eqref{Slightly compressible strain energy} was originally conceived of for hyperelastic solids, it can be readily incorporated in the MQLV constitutive equation~\eqref{Cauchy viscoelastic stress} through the elastic Cauchy stress tensor $\bm{T}^{\mathrm{e}}$ or, equivalently, through the elastic second Piola--Kirchhoff tensor $\bm{\Pi}^{\mathrm{e}}$. By substituting~\eqref{Slightly compressible strain energy} into~\eqref{Second Piola-Kirchhoff stress tensors compressible} and applying the chain rule, we obtain:
\begin{equation}
	\left\{
	\begin{aligned}
	&\bm{\Pi}^{\mathrm{e}}_{\mathrm{H}}=J^{1 \slash 3}U^{\prime}(J)@@\overline{\bm{C}}^{-1}, 
        \\	
	&\bm{\Pi}^{\mathrm{e}}_{\mathrm{D}}=\frac{2}{3}J^{-2 \slash 3}\left[3\psi_1\bm{I}-\left(\bar{I}_1\psi_1-\bar{I}_2\psi_2\right)\overline{\bm{C}}^{-1}-\psi_2\overline{\bm{C}}^{-2}
    \right],
	\end{aligned}
	\label{Second Piola-Kirchhoff stress tensors slightly compressible}
    \right.
\end{equation}
where $\psi_i=\partial \psi \slash \partial \bar{I}_i$ for $i=1,2,3$. Substitution of~\eqref{Second Piola-Kirchhoff stress tensors slightly compressible} into~\eqref{Cauchy viscoelastic stress} then yields the MQLV constitutive equation for an isotropic slightly compressible viscoelastic solid with instantaneous hyperelastic response.

In this work, we further specialise this MQLV formulation by specifying the following slightly compressible Mooney--Rivlin strain energy function~\cite{small24,wang21,FEBioUserManualUncoupledMR}:
\begin{equation}
    W=c_1(\bar{I}_1-3)+c_2(\bar{I}_2-3)+\frac{1}{2}\kappa_0(\ln J)^2,
    \label{Nearly-incompressible MR strain energy}
\end{equation}
corresponding to $\psi=c_1(\bar{I}_1-3)+c_2(\bar{I}_2-3)$ and $U=\kappa_0(\ln J)^2 \slash 2$ in~\eqref{Slightly compressible strain energy}. We note in passing that a myriad of alternative forms for the volumetric term $U$ have been proposed in the literature (see, for example,~\cite{pelliciari23,anand20,horgan07,doll00}), which differ primarily in the associated mechanical response for large and small volume dilations~\cite{anand20}. The stress relaxation behaviour of this slightly compressible model will be considered later.

%%%%%%%%%%%%%%%%%%%%%%%%%%%%%%%%%%%%%%%%%%%%%%%%%%%%%%
\subsection{Incompressible limit}
\label{sec: incompressible mqlv}
To conclude this section, we examine the incompressible limit, following the approach of~\cite{righi21}. In this limit, the volume ratio tends to unity ($J(t) \to 1$), while the bulk modulus tends to infinity ($\kappa(t) \to \infty$), and so~\eqref{Cauchy viscoelastic stress} becomes:   
\begin{equation}
    \bm{T}(t)=\bm{F}(t)\left(\bm{\Pi}^{\mathrm{e}}_{\mathrm{D}}(t)+\frac{1}{\mu_0}\int_{0}^{t}\mu^{\prime}(t-s)\bm{\Pi}^{\mathrm{e}}_{\mathrm{D}}(s)\dd{s}\right)\bm{F}(t)^{\transpose}-p(t)\bm{I},    
    \label{Cauchy viscoelastic stress incompressible form}
\end{equation}
where now, from~\eqref{MQLV Piola transforms and hyd/dev splits}, we have:
\begin{equation}
    \bm{\Pi}^{\mathrm{e}}_{\mathrm{D}}=2\left[W_1\bm{I}+\frac{1}{3}\left(I_2W_2-I_1W_1\right)\bm{C}^{-1}-W_2\bm{C}^{-2}\right] 
    \label{PiD incompressible form}
\end{equation}
and the Lagrange multiplier of incompressibility is given by:
\begin{equation}
    p(t)=-\displaystyle \lim_{\substack{J(t) \to 1, \\[1pt] \kappa(t) \to \infty}} J(t)^{-1}\bm{F}(t)\left(\bm{\Pi}^{\mathrm{e}}_{\mathrm{H}}(t)+\frac{1}{\kappa_0}\int_{0}^{t}\kappa^{\prime}(t-s)\bm{\Pi}^{\mathrm{e}}_{\mathrm{H}}(s)\dd{s}\right)\bm{F}(t)^{\transpose}.
    \label{Cauchy viscoelastic stress Lagrange multiplier}
\end{equation}
We note that in the incompressible limit, the relaxation function for the bulk modulus $\kappa(t)$ is not prescribed explicitly, as in~\eqref{Prony series}. Rather, it is absorbed into the Lagrange multiplier $p(t)$ in~\eqref{Cauchy viscoelastic stress incompressible form}, which itself is introduced to enforce the incompressibility constraint. Accordingly, $p(t)$ is determined by solving the equation of motion subject to appropriate boundary conditions. Under the common assumptions that the deformation is slow enough that inertial effects can be neglected (the quasi-static assumption) and external body forces are negligible compared to applied surface forces, the equation of motion reduces to~\cite{destrade25, anand20, atkin13, holzapfel00, ogden97}:
\begin{equation}
    \mathrm{div}\,\bm{T}=\bm{0},
    \label{Momentum balance equation}
\end{equation}
where $\mathrm{div}$ denotes the divergence operator in the deformed configuration.

In this work, we further specialise this incompressible MQLV model by specifying the following incompressible Mooney--Rivlin strain energy function~\cite{small25,balbi19,destrade15,rashid13}:
\begin{equation}
    W=c_1(I_1-3)+c_2(I_2-3).
    \label{Incompressible MR strain energy}
\end{equation}

In what follows, we use both the incompressible and slightly compressible Mooney--Rivlin MQLV models formulated in this section, with Prony series relaxation functions given by~\eqref{Prony series}, to analyse their stress relaxation response under four different deformation modes: isochoric and nearly-isochoric simple shear, followed by isochoric and nearly-isochoric torsion.

%%%%%%%%%%%%%%%%%%%%%%%%%%%%%%%%%%%%%%%%%%%%%%%%%%%%%%
\section{Simple shear of a solid cuboid}
\label{sec: simple shear}

In this section, we consider both the isochoric and nearly-isochoric simple shear of a solid cuboid and derive analytical expressions for the corresponding shear and normal stresses for the incompressible, compressible and slightly compressible MQLV models formulated in Section~\ref{sec: mqlv}.

%%%%%%%%%%%%%%%%%%%%%%%%%%%%%%%%%%%%%%%%%%%%%%%%%%%%%%
\subsection{Isochoric simple shear}
\label{sec: isochoric simple shear}

\begin{figure}[h!]
    {\centerline{\includegraphics[scale=0.5]{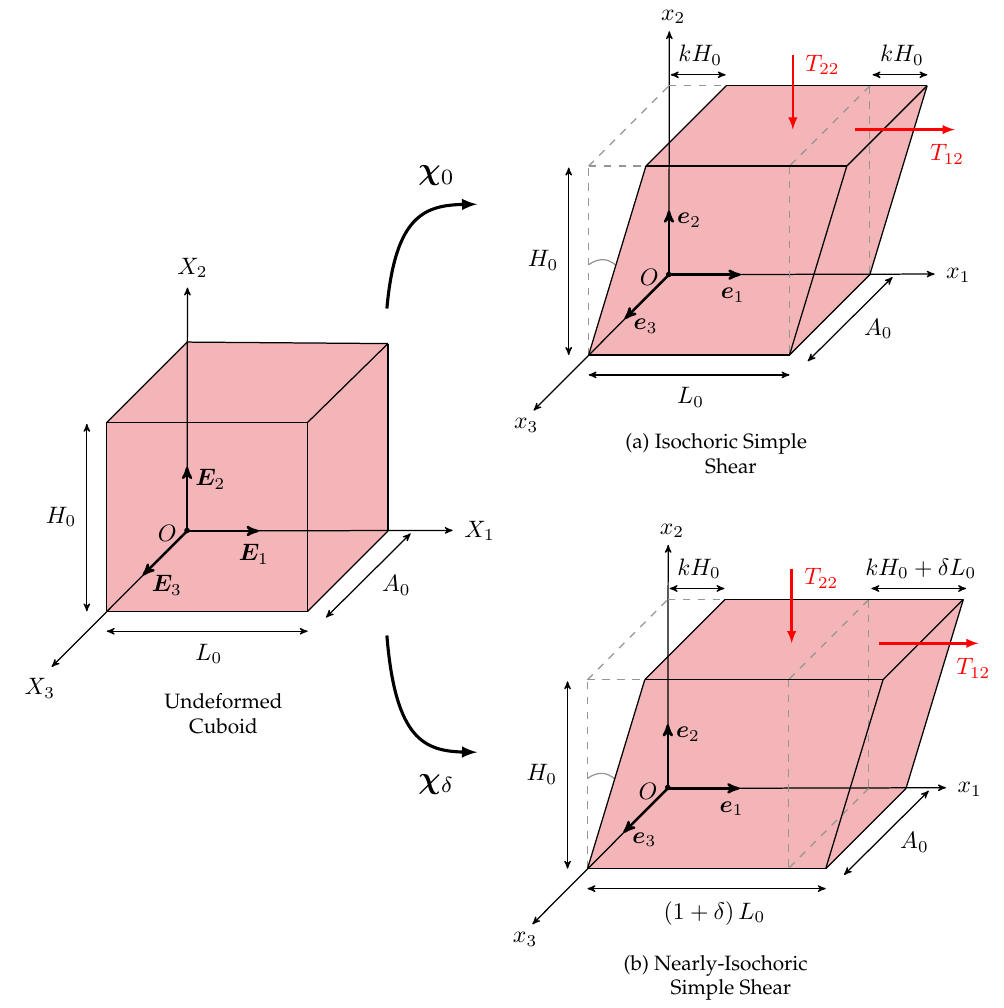}}}
    \caption{Deformation of a cuboid under (a)~isochoric simple shear $\bm{\chi}_0$ and (b)~nearly-isochoric simple shear $\bm{\chi}_{\delta}$. In both cases, shear and normal tractions must be applied to the top face of the cuboid to maintain the deformation.}
    \label{fig: simple shear deformations}
\end{figure}

In classical simple shear, a cuboid is deformed into a parallelepiped by displacing the top face relative to the fixed bottom face through the application of both shear and normal tractions, without changing the volume or dimensions of the cuboid (see Figure~\ref{fig: simple shear deformations}(a)). This deformation, subsequently referred to as isochoric simple shear, can be written as:
\begin{equation}
\left\{
    \begin{aligned}
        &x_1=X_1+k(t)X_2, 
        \\
        &x_2=X_2, 
        \\
        &x_3=X_3,
    \end{aligned}
    \label{Isochoric simple shear}
\right.
\end{equation}
where $(X_1,X_2,X_3)$ and $(x_1,x_2,x_3)$ denote the Cartesian coordinates of a material point before and after deformation, respectively, and $k(t)$ is the amount of shear. By introducing the fixed Cartesian bases $\{\bm{E}_{1},\bm{E}_{2},\bm{E}_{3}\}$ and $\{\bm{e}_{1},\bm{e}_{2},\bm{e}_{3}\}$ for the undeformed and deformed configurations, respectively, we can write the deformation gradient $\bm{F}=F_{aA}\,\bm{e}_a\otimes\bm{E}_A$ associated with the deformation~\eqref{Isochoric simple shear} as follows:
\begin{equation}
    \bm{F}(t)=\left(\begin{array}{ccc}
                1 & k(t) & 0 \\
                0 & 1 & 0 \\
                0 & 0 & 1
        \end{array}\right),
        \label{Deformation gradient isochoric simple shear}
\end{equation}
with $J=\mathrm{det}\,\bm{F}=1$. The right Cauchy--Green deformation tensor $\bm{C}=\bm{F}^{\transpose}\bm{F}$ is then given by:
\begin{equation}
    \bm{C}(t)=\left(\begin{array}{ccc}
                1 & k(t) & 0 \\[1pt]
                k(t) & k(t)^2+1 & 0 \\[1pt]
                0 & 0 & 1
        \end{array}\right),
        \label{C isochoric simple shear}
\end{equation}
from which we can easily compute the corresponding principal invariants~\eqref{Principal invariants}:
\begin{equation}
    I_{1}=k^2+3, \qquad
    I_{2}=k^2+3 \qquad \text{and} \qquad  
    I_{3}=1.
    \label{Principal invariants isochoric simple shear}
\end{equation}

Since the deformation is homogeneous, the equation of motion~\eqref{Momentum balance equation} is automatically satisfied for both the incompressible and slightly compressible MQLV models considered in this paper. Maintaining the deformation nevertheless requires the application of both shear and normal tractions on the top face of the cuboid. In this section, we derive analytical expressions for the corresponding stresses for each model.

Before proceeding, we note in passing that, in addition to the shear and normal tractions mentioned above, surface tractions are also required on the slanted faces of the deformed cuboid. In reality, however, such tractions are never applied in simple shear experiments, causing the slanted faces to bend and the deformation to become inhomogeneous, resulting in large variations in the distribution of the stresses~\cite{destrade23}. These undesirable effects are minimised in practice by using sufficiently thin cuboids, with the length and width equal and both at least four times greater than the height~\cite{destrade23,destrade15,rashid13}.

%%%%%%%%%%%%%%%%%%%%%%%%%%%%%%%%%%%%%%%%%%%%%%%%%%%%%%
\subsubsection{Incompressible model}
\label{sec: isochoric simple shear incompressible}

For the incompressible Mooney--Rivlin MQLV model, the partial derivatives of the strain energy function~\eqref{Incompressible MR strain energy} are:
\begin{equation}
W_1=c_1 \qquad \text{and} \qquad W_2=c_2.
\label{Incompressible MR strain energy derivatives}
\end{equation}
Substituting these derivatives into~\eqref{PiD incompressible form}, together with~\eqref{C isochoric simple shear} and~\eqref{Principal invariants isochoric simple shear}, we obtain the corresponding expression for $\bm{\Pi}^{\mathrm{e}}_{\mathrm{D}}$. The MQLV constitutive equation~\eqref{Cauchy viscoelastic stress incompressible form} then gives the following non-zero components of the Cauchy viscoelastic stress tensor:

\begin{alignat}{2}
    &T_{11}(t)&&=-p(t)+\frac{2}{3}\left(2c_1+c_2\right)k(t)^2-2\mu_0k(t)\mathrm{DInt}_1(t)
    +\frac{2}{3}\Big[\left(c_1+2c_2\right)k(t)^2 \nonumber \\ 
    &&&+4c_1+5c_2\Big]\mathrm{DInt}_2(t)-\frac{2}{3}\left(c_1+2c_2\right)\left(2k(t)\mathrm{DInt}_3(t)-\mathrm{DInt}_4(t)\right),  
    \label{T11 isochoric simple shear incompressible} \\[5pt] 
    &T_{12}(t)&&=\mu_0\left(k(t)-\mathrm{DInt}_1(t)\right)+\frac{2}{3}\left(c_1+2c_2\right)\left(k(t)\mathrm{DInt}_2(t)-\mathrm{DInt}_3(t)\right), 
    \label{T12 isochoric simple shear incompressible} \\[5pt]
    &T_{22}(t)&&=-p(t)-\frac{2}{3}\left(c_1+2c_2\right)(k(t)^2-\mathrm{DInt}_2(t)), 
    \label{T22 isochoric simple shear incompressible} \\[5pt]
    &T_{33}(t)&&=-p(t)-\frac{2}{3}\left(c_1-c_2\right)(k(t)^2-\mathrm{DInt}_2(t)),
    \label{T33 isochoric simple shear incompressible}
\end{alignat}

where we have introduced the compact notation in~\eqref{Deviatoric integrals shortand notation} for the integral terms $\mathrm{DInt}_i(t)$.

As is standard for the isochoric simple shear of an incompressible solid, we assume that the out-of-plane surface of the cuboid is traction-free throughout the deformation, so that~\cite{balbi18,depascalis15}:
\begin{equation}
    T_{33}(t)=0.
    \label{Isochoric simple shear BC}
\end{equation}
By imposing this boundary condition in~\eqref{T33 isochoric simple shear incompressible}, we obtain the Lagrange multiplier of incompressibility:
\begin{equation}
    p(t)=-\frac{2}{3}\left(c_1-c_2\right)(k(t)^2-\mathrm{DInt}_2(t)),
    \label{Isochoric simple shear Lagrange mutiplier}
\end{equation}
which, upon substitution into~\eqref{T11 isochoric simple shear incompressible}--\eqref{T22 isochoric simple shear incompressible}, yields the fully determined stress components: 
\begin{alignat}{2}
    &T_{11}(t)&&=2c_1k(t)^2-2\mu_0k(t)\mathrm{DInt}_1(t)+\frac{2}{3}\left(c_1+2c_2\right)\left(3+k(t)^2\right)\mathrm{DInt}_2(t) \\ 
    &&&-\frac{2}{3}\left(c_1+2c_2\right)\left(2k(t)\mathrm{DInt}_3(t)-\mathrm{DInt}_4(t)\right), 
    \label{T11 isochoric simple shear incompressible final} \\[5pt] 
    &T_{12}(t)&&=\mu_0\left(k(t)-\mathrm{DInt}_1(t)\right)+\frac{2}{3}\left(c_1+2c_2\right)\left(k(t)\mathrm{DInt}_2(t)-\mathrm{DInt}_3(t)\right), 
    \label{T12 isochoric simple shear incompressible final} \\[5pt]
    &T_{22}(t)&&=-2c_2(k(t)^2-\mathrm{DInt}_2(t)).
    \label{T22 isochoric simple shear incompressible final}
\end{alignat}

%%%%%%%%%%%%%%%%%%%%%%%%%%%%%%%%%%%%%%%%%%%%%%%%%%%%%%
\subsubsection{Slightly compressible model}
\label{sec: isochoric simple shear slightly compressible}

For the slightly compressible Mooney--Rivlin MQLV model, the partial derivatives of the strain energy function~\eqref{Nearly-incompressible MR strain energy} are:
\begin{equation}
    W_1=c_1, \qquad
    W_2=c_2 \qquad \text{and} \qquad  
    W_3=\frac{1}{8}\kappa_0J^{-1 \slash 2}\ln J.
    \label{Slightly compressible MR strain energy derivatives}
\end{equation}
Since the deformation~\eqref{Isochoric simple shear} is isochoric (i.e.~$J=1$), the modified right Cauchy--Green deformation tensor $\overline{\bm{C}}$ and its associated invariants $\bar{I}_1$ and $\bar{I}_2$ reduce to their unmodified counterparts given by~\eqref{C isochoric simple shear} and~\eqref{Principal invariants isochoric simple shear}, respectively. The corresponding expressions for $\bm{\Pi}^{\mathrm{e}}_{\mathrm{H}}$ and $\bm{\Pi}^{\mathrm{e}}_{\mathrm{D}}$ then follow from~\eqref{Second Piola-Kirchhoff stress tensors slightly compressible}, and substitution into the MQLV constitutive equation~\eqref{Cauchy viscoelastic stress} yields the following non-zero components of the Cauchy stress tensor:
\begin{alignat}{2}
    &T_{11}(t)&&=\frac{2}{3}\left(2c_1+c_2\right)k(t)^2-2\mu_0k(t)\mathrm{DInt}_1(t)
    +\frac{2}{3}\left[\left(c_1+2c_2\right)k(t)^2+4c_1+5c_2\right]\mathrm{DInt}_2(t)\nonumber \\
    &&&-\frac{2}{3}\left(c_1+2c_2\right)\left(2k(t)\mathrm{DInt}_3(t)
    -\mathrm{DInt}_4(t)\right), 
    \label{T11 isochoric simple shear slightly compressible} \\[5pt]
    &T_{12}(t)&&=\mu_0(k(t)-\mathrm{DInt}_1(t))+\frac{2}{3}\left(c_1+2c_2\right)\left(k(t)\mathrm{DInt}_2(t)-\mathrm{DInt}_3(t)\right), 
    \label{T12 isochoric simple shear slightly compressible} \\[5pt]
    &T_{22}(t)&&=-\frac{2}{3}\left(c_1+2c_2\right)(k(t)^2-\mathrm{DInt}_2(t)), 
    \label{T22 isochoric simple shear slightly compressible} \\[5pt]
    &T_{33}(t)&&=-\frac{2}{3}\left(c_1-c_2\right)(k(t)^2-\mathrm{DInt}_2(t)),
    \label{T33 isochoric simple shear slightly compressible}
\end{alignat}
which coincide with the stress components~\eqref{T11 isochoric simple shear incompressible}--\eqref{T33 isochoric simple shear incompressible} in the absence of the Lagrange multiplier of incompressibility.

%%%%%%%%%%%%%%%%%%%%%%%%%%%%%%%%%%%%%%%%%%%%%%%%%%%%%%
\subsection{Nearly-isochoric simple shear}
\label{sec: nearly isochoric simple shear}

The isochoric deformation~\eqref{Isochoric simple shear} assumes that simple shear is accompanied by no change in volume. This assumption is routinely adopted for soft solids, which are typically modelled as incompressible. In reality, however, all materials are to some extent compressible, so experimental simple shear is invariably accompanied by an infinitesimal change in volume. Following Destrade \textit{et al.}~\cite{destrade15}, we model this infinitesimal volume change with the following simple generalisation of the isochoric deformation~\eqref{Isochoric simple shear}:
\begin{equation}
\left\{
    \begin{aligned}
        &x_1=\left(1+\delta\right)X_1+k(t)X_2, 
        \\
        &x_2=X_2, 
        \\
        &x_3=X_3,
    \end{aligned}
    \label{Nearly isochoric simple shear}
\right.
\end{equation}
which corresponds to simple shear of amount $k(t)$ superposed on uniaxial extension of amount $\delta$ in the shear direction (see Figure~\ref{fig: simple shear deformations}(b)). The associated deformation gradient takes the form: 
\begin{equation}
    \bm{F}(t)=\left(\begin{array}{ccc}
                1+\delta & k(t) & 0 \\
                0 & 1 & 0 \\
                0 & 0 & 1
        \end{array}\right),
        \label{Deformation gradient nearly isochoric simple shear}
\end{equation}
with $J=1+\delta$; hence, $\delta$ can alternatively be interpreted as the relative volume change between the undeformed and deformed configurations. The corresponding modified right Cauchy--Green deformation tensor $\overline{\bm{C}}=J^{-2 \slash 3}\bm{F}^{\transpose}\bm{F}$ reads:
\begin{equation}
    \overline{\bm{C}}(t)= \left(1+\delta\right)^{-2 \slash 3}\left(\begin{array}{ccc}
                \left(1+\delta\right)^{2} & \left(1+\delta\right)k(t) & 0 \\[1pt]
                \left(1+\delta\right)k(t) & k(t)^2+1 & 0 \\[1pt]
                0 & 0 & 1
        \end{array}\right)
        \label{C nearly isochoric simple shear}
\end{equation}
and has modified principal invariants $\bar{I}_{1}=\mathrm{tr}\,\overline{\bm{C}}$ and $\bar{I}_{2}=\mathrm{tr}\,\overline{\bm{C}}^{-1}$ given by:
\begin{equation}
    \bar{I}_{1}=\left(1+\delta\right)^{-2 \slash 3}\left[k(t)^2+2+\left(1+\delta\right)^2\right] \quad \text{and} \quad  
    \bar{I}_{2}=\left(1+\delta\right)^{-4 \slash 3}\left[k(t)^2+3+2\delta^2+4\delta\right].
    \label{Principal invariants nearly isochoric simple shear}
\end{equation}

We note that this modelling assumption results in a positive volume change during shearing, as demonstrated in~\cite{destrade15}. Moreover, as the volume change associated with the deformation~\eqref{Nearly isochoric simple shear} is assumed to be small (i.e. $0 \leq \delta \ll 1$), we use the slightly compressible MQLV model to derive analytical expressions for the corresponding stresses. In this case, the partial derivatives of the strain energy function~\eqref{Nearly-incompressible MR strain energy} are given by~\eqref{Slightly compressible MR strain energy derivatives}, which can be combined with~\eqref{C nearly isochoric simple shear} and~\eqref{Principal invariants nearly isochoric simple shear} to obtain the corresponding expressions for $\bm{\Pi}^{\mathrm{e}}_{\mathrm{H}}$ and $\bm{\Pi}^{\mathrm{e}}_{\mathrm{D}}$ in~\eqref{Second Piola-Kirchhoff stress tensors slightly compressible}. The MQLV constitutive equation~\eqref{Cauchy viscoelastic stress} then gives the following non-zero components of the Cauchy stress tensor up to first-order in $\delta$:
\begin{alignat}{2}
    &T_{11}(t)&&=\frac{2}{3}\left(2c_1+c_2\right)k(t)^2-2\mu_0k(t)\mathrm{DInt}_1(t)+\frac{2}{3}\left[\left(c_1+2c_2\right)k(t)^2+4c_1+5c_2\right]\mathrm{DInt}_2(t)
    \nonumber \\
    &&&-\frac{2}{3}\left(c_1+2c_2\right)\left(2k(t)\mathrm{DInt}_3(t)-\mathrm{DInt}_4(t)\right)-\frac{1}{9}\delta \Big[2\left(10c_1+7c_2\right)k(t)^2-12\mu_0-9\kappa_0\nonumber \\
    &&&-6\mu_{0}(k(t)^2-2)\mathrm{DInt}_0(t)-12\left(3c_1+5c_2\right)k(t)\mathrm{DInt}_1(t)\nonumber \\
    &&&+2(14c_1+29c_2+\left(5c_1+14c_2\right)k(t)^2)\mathrm{DInt}_2(t)-(5c_1+14c_2)(4k(t)\mathrm{DInt}_3(t)-\mathrm{DInt}_4(t))\nonumber \\
    &&&+9\kappa_0(k(t)^2+1)\mathrm{HInt}_0(t)-9\kappa_0(2k(t)\mathrm{HInt}_1(t)-\mathrm{HInt}_2(t))\Big], 
\label{T11 nearly isochoric simple shear} \\[5pt]
    &T_{12}(t)&&=\mu_0(k(t)-\mathrm{DInt}_1(t))+\frac{2}{3}\left(c_1+2c_2\right)(k(t)\mathrm{DInt}_2(t)-\mathrm{DInt}_3(t)) \nonumber \\
    &&&-\frac{1}{9}\delta\Big[6\left(5c_1+7c_2\right)
    -6\mu_0k(t)\mathrm{DInt}_0(t)
    -6\left(3c_1+5c_2\right)\mathrm{DInt}_1(t)\nonumber \\
    &&&
    +2\left(5c_1+14c_2\right)(k(t)\mathrm{DInt}_2(t) -\mathrm{DInt}_3(t))+9\kappa_0(k(t)\mathrm{HInt}_0(t)-\mathrm{HInt}_1(t))\Big], 
\label{T12 nearly isochoric simple shear} \\[5pt]
    &T_{22}(t)&&=-\frac{2}{3}\left(c_1+2c_2\right)(k(t)^2-\mathrm{DInt}_2(t))-\frac{1}{9}\delta\Big[6\mu_0-9\kappa_0-6\mu_0\mathrm{DInt}_0(t) \nonumber \\
    &&&-2\left(5c_1+14c_2\right)(k(t)^2-\mathrm{DInt}_2(t))+9\kappa_0\mathrm{HInt}_0(t)\Big], 
\label{T22 nearly isochoric simple shear} \\[5pt]
    &T_{33}(t)&&=-\frac{2}{3}\left(c_1-c_2\right)(k(t)^2-\mathrm{DInt}_2(t))-\frac{1}{9}\delta\Big[6\mu_0-9\kappa_0-6\mu_0\mathrm{DInt}_0(t) \nonumber \\
    &&&-2\left(5c_1-7c_2\right)(k(t)^2-\mathrm{DInt}_2(t))+9\kappa_0\mathrm{HInt}_0(t)\Big],
\label{T33 nearly isochoric simple shear}
\end{alignat}
where the terms $\mathrm{HInt}_i(t)$ are defined in~\eqref{Hydrostatic integrals shortand notation}.
The corresponding stress components for the isochoric slightly compressible MQLV model, given by~\eqref{T11 isochoric simple shear slightly compressible}--\eqref{T33 isochoric simple shear slightly compressible}, are recovered as a special case when $\delta=0$. 

%%%%%%%%%%%%%%%%%%%%%%%%%%%%%%%%%%%%%%%%%%%%%%%%%%%%%%

\section{Torsion of a solid cylinder}
\label{sec:theotorsion}

\begin{figure}[t!]
    {\centerline{\includegraphics[scale=0.5]{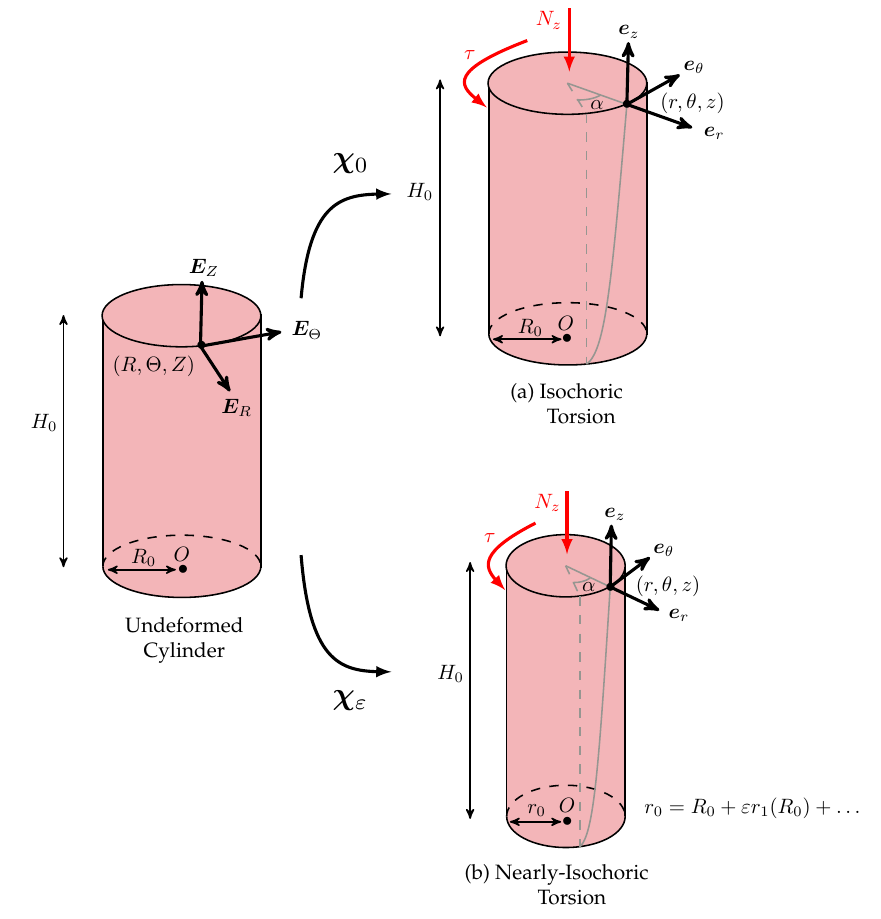}}}
    \caption{Deformation of a cylinder under (a)~isochoric torsion $\bm{\chi}_0$ and (b)~nearly-isochoric torsion $\bm{\chi}_{\varepsilon}$. In both cases, torque and axial force must be applied to the top face of the cylinder to maintain the deformation.}
    \label{fig: torsion deformations}
\end{figure}

In this section, we derive the equations for the torsion of a solid cylinder. We consider both the incompressible and the slightly compressible cases. The deformation can be written as follows:
\begin{equation}
\left\{
\begin{aligned}
&r = r(R,t), \\
&\theta = \Theta + \phi(t)Z, \\
&z = Z,
\end{aligned}
\right.
\label{eq:torsion-def}
\end{equation}
where the twist $\phi=\alpha \slash H_0$ is the angle of rotation per unit height (see Figure~\ref{fig: torsion deformations}). By introducing the cylindrical bases $\{\bm{E}_{R},\bm{E}_{\Theta},\bm{E}_{Z}\}$ and $\{\bm{e}_{r},\bm{e}_{\theta},\bm{e}_{z}\}$ for the undeformed and deformed configurations, respectively, we can write the deformation gradient $\bm{F}=F_{aA}\,\bm{e}_a\otimes\bm{E}_A$ associated with the deformation~\eqref{eq:torsion-def} as follows: 
\begin{equation}
\bm{F}
=
\begin{pmatrix}
r^{\prime} & 0 & 0 \\
0 & r \slash R & r\phi  \\
0 & 0 & 1
\end{pmatrix},
\label{eq:Ftorsion}
\end{equation}
where $r^{\prime}=\mathrm{d}r \slash \mathrm{d}R$ and $J=rr^{\prime} \slash R$.
For all $t \geq 0$, the motion of a solid cylinder in torsion is governed by the following boundary value problem~\cite{small25,righi21}:
\begin{equation}
\left\{
\begin{aligned}
&T_{rr}'+(T_{rr}-T_{\theta\theta})\dfrac{r'(R)}{r(R)}=0,\\
&T_{rr}(R_0) = 0,
\end{aligned}
\right.
\label{eq:divtorincomp}
\end{equation}
which follows from the equation of motion~\eqref{Momentum balance equation}. Here, the radial $T_{rr}$ and circumferential $T_{\theta\theta}$ components of the Cauchy stress tensor are given by the corresponding constitutive equation, either in the incompressible form~\eqref{Cauchy viscoelastic stress incompressible form} or in the compressible form~\eqref{Cauchy viscoelastic stress}. In addition, $r_0=r(R_0)$ is the outer radius of the cylinder in the deformed state, whereas $R_0$ is the initial undeformed outer radius (see Figure~\ref{fig: torsion deformations}). The governing equation is complemented by the zero-traction boundary condition, requiring the radial component of the Cauchy stress tensor $T_{rr}$ to be zero at $R=R_0$ for $t\geq0$. Finally, the resulting torque $\tau$ and axial force $N_z$ required to twist the cylinder are given by~\cite{small24,polignone91}:
\begin{equation}
\tau = 2\pi \int_{0}^{R_0} T_{\theta z}(R)\,r(R)^2\,r^{\prime}(R)\dd{R} \qquad \text{and}\qquad N_z = 2\pi \int_{0}^{R_0} T_{zz}(r)\,r(R)\,r^{\prime}(R)\dd{R}.
\label{eq:torqueforce}
\end{equation}

In the next section, these formulas will be used to compute the resultant torque and axial force predicted by the MQLV model for both incompressible and slightly compressible cylinders in torsion.

%%%%%%%%%%%%%%%%%%%%%%%%%%%%%%%%%%%%%%%%%%%%%%%%%%%%%%

\subsection{Torsion of an incompressible cylinder}
\label{sec:torsion of an incompressible cylinder}

We begin by considering the incompressible case. For an incompressible material, the deformation~\eqref{eq:torsion-def} is isochoric (see Figure~\ref{fig: torsion deformations}(a)). Therefore, the deformation gradient satisfies $\mathrm{det}\,\bm{F}=1$, which leads to $r=R$ for $t\geq 0$. We use the Mooney--Rivlin strain energy function for an incompressible material~\eqref{Incompressible MR strain energy}.
Combining~\eqref{Incompressible MR strain energy} with~\eqref{PiD incompressible form} and substituting into the constitutive equation~\eqref{Cauchy viscoelastic stress incompressible form}, we obtain the non-zero components of the Cauchy stress tensor:
\begin{alignat}{2}
&T_{rr}(t) &&=-p(t)- \frac{2}{3}(c_1 - c_2) r^{2}\bigl(\phi(t)^{2}-\mathrm{DInt}_2(t)\bigr),
\label{Trr} \\[5pt]
&T_{\theta\theta}(t) &&=-p(t)+ \frac{2}{3}\Bigl[ (2c_1 + c_2)\phi(t)^2- 3\mu_0\mathrm{DInt}_1(t)\phi(t)
 +(4c_1 + 5c_2)\mathrm{DInt}_2(t)\Bigr] r^2 \nonumber \\
&&&+ \frac{2}{3}(c_1 + 2c_2)\Bigl(\mathrm{DInt}_2\phi(t)^2- 2\mathrm{DInt}_3(t)\phi(t)+\mathrm{DInt}_4(t)\Bigr) r^4,
\label{Ttt}  \\[5pt]
&T_{\theta z}(t) &&=
\mu_0r\bigl(\phi(t)-\mathrm{DInt}_1(t)\bigr)+\frac{2}{3}(c_1 + 2c_2) r^3
\bigl(\mathrm{DInt}_3(t)-\mathrm{DInt}_2(t)\phi(t)\bigr)
,
\label{Ttz}  \\[5pt]
&T_{zz}(t) &&=-p(t)
- \frac{2}{3}(c_1 + 2c_2) r^{2}
\bigl(\phi(t)^{2}-\mathrm{DInt}_2(t)\bigr),
 \label{Tzz}
\end{alignat}
where, the integrals $\mathrm{DInt}_i(t)$ are defined in~\eqref{Deviatoric integrals shortand notation}.
By using~\eqref{Trr} and~\eqref{Ttt}, we can now solve the governing problem in~\eqref{eq:divtorincomp} for $p(t)$ to obtain:
\begin{align}
p(t)&=- \frac{1}{6}
 \Bigl[
2(5c_1 - 2c_2) r^{2}
- 6c_1 R_0^{2}
\Bigr]\phi(t)^{2}+\mu_0(r^{2} - R_0^{2})\phi(t)\mathrm{DInt}_1(t) \nonumber  \\ 
&- \Bigl[
\frac{1}{3}(c_1 + 8c_2) r^{2}
- (c_1 + 2c_2) R_0^{2}
\Bigr]\mathrm{DInt}_2(t)
 \\
&-\frac{1}{6}(c_1 + 2c_2)(r^{4} - R_0^{4})\Bigl(
 \mathrm{DInt}_2(t)\phi(t)^{2}
 - 2\mathrm{DInt}_3(t)\phi(t)
 +\mathrm{DInt}_4(t)
\Bigr).
\label{eq:p}
\end{align}

Now that all the stress components in~\eqref{Trr}--\eqref{Tzz} are fully determined, we can use~\eqref{Ttz} and~\eqref{Tzz} to compute the torque and axial force in~\eqref{eq:torqueforce} as follows:
\begin{align}
\tau(t) &=
\frac{\pi}{2}\mu_0
\bigl(\phi(t) - \mathrm{DInt}_1(t)\bigr) R_0^{4}
+ \frac{2\pi}{9}(c_1 + 2c_2)
\bigl(\mathrm{DInt}_2(t)\phi(t) - \mathrm{DInt}_3(t)\bigr) R_0^{6}, 
\label{eq:tau} \\[5pt]  
N_z(t) &=
-\frac{\pi}{2}
\Bigl[
(c_1 + 2c_2)\phi(t)^2
- \mu_0\mathrm{DInt}_1(t)\phi(t)
+c_1\mathrm{DInt}_2(t)
\Bigr] R_0^{4} \nonumber \\
&-\frac{\pi}{9}(c_1 + 2c_2)
\Bigl(
\mathrm{DInt}_2(t)\phi(t)^2
- 2\mathrm{DInt}_3(t)\phi(t)
+\mathrm{DInt}_4(t)
\Bigr) R_0^{6}.
\label{eq:Fz}
\end{align}

In the next section, we consider the deformation of a slightly compressible solid cylinder in torsion.

%%%%%%%%%%%%%%%%%%%%%%%%%%%%%%%%%%%%%%%%%%%%%%%%%%%%%%

\subsection{Torsion of a slightly compressible cylinder}
\label{sec:torsion of a slightly compressible cylinder}

Here, we adapt the perturbation method proposed by Levinson~\cite{levinson1972} and used by Small \textit{et al.}~\cite{small24} to model the torsion of a hyperelastic cylinder. As shown in Figure~\ref{fig: torsion deformations}(b), we start by perturbing the isochoric deformation~\eqref{eq:torsion-def}, for which $r=R$, by allowing for a small volume change in the radial direction and adopting the following ansatz for the radial deformation:
\begin{equation}
r = R + \varepsilon  r_1(R) + \varepsilon^2 r_2(R)+ O(\varepsilon^3),
\label{eq:rEPS}
\end{equation}
where $\varepsilon=\mu_0 \slash \kappa_0 \ll 1$ is a small parameter that describes the degree of slight compressibility, with $\varepsilon=0$ corresponding to incompressible behaviour. Hence, the volume ratio $J=rr^{\prime} \slash R$ expanded up to second-order in $\varepsilon$ reads:
\begin{equation}
J = 1 + \varepsilon  J_1(R) + \varepsilon^2 J_2(R)+ O(\varepsilon^3), 
\label{eq:JEPS}
\end{equation}
where:
\begin{equation}
	\left\{
	\begin{aligned}
	& J_1=r_1^{\prime}(R)+\dfrac{r_1(R)}{R}, 
    \label{eq:J1} \\
    & J_2=r_2^{\prime}(R)+\frac{ r_2(R)+r_1(R) r_1^{\prime}(R)}{R}.
	\end{aligned}
    \right.
\end{equation}

We use the slightly compressible Mooney--Rivlin strain energy function~\eqref{Nearly-incompressible MR strain energy} and compute the tensors $\bm{\Pi}^{\mathrm{e}}_{\mathrm{H}}$ and $\bm{\Pi}^{\mathrm{e}}_{\mathrm{D}}$ in~\eqref{Second Piola-Kirchhoff stress tensors slightly compressible}. We then insert these expressions into the compressible MQLV constitutive equation~\eqref{Cauchy viscoelastic stress}, expand the components of the Cauchy stress tensor $\bm{T}(t)$ up to first-order in $\varepsilon$. The resulting components of the Cauchy stress tensor display the following form:
\begin{equation}
    T_{ij}(t)=T_{ij}^0(r_1(R),J_1(R),R,t)+\varepsilon @T_{ij}^1(r_1(R),J_1(R),J_2(R),R,t)\quad\text{for}\quad i,j \in \{r,\theta,z\}.
    \label{eq:Tijform}
\end{equation}
Upon substituting them into the governing equation~\eqref{eq:divtorincomp}, we obtain the governing problem at order zero as follows:
\begin{equation}
	\left\{
	\begin{aligned}
	& r_1^{\prime \prime}(R)+ \left(\dfrac{1}{R} - \dfrac{A_2}{A_0} R\right) r_1^{\prime}(R)- \left(\dfrac{A_2}{A_0} + \frac{1}{R^{2}}\right) r_1(R)=\dfrac{A_1R+ A_3R^3}{A_0}, \\
    & r_1^{\prime}(R_0)+\frac{r_1(R_0)}{R_0}=C_0 R_0^2,
	\end{aligned}
    \right.
    \label{eq:gov0}
\end{equation}
where we have introduced the coefficients $A_i(t)$, which are functions of $t$ only and are listed in \eqref{app:A0}--\eqref{app:A3}, together with the constant term $C_0$ in \eqref{app:C0}.
An explicit analytical solution of~\eqref{eq:gov0} can be found by using the Frobenius method~\cite{Kreyszig11}. Seeking solutions of the form:
\begin{equation}
    r_1(R)=R^s\sum_{n=0}^{\infty}a_n R^n,
\end{equation}
we find that the indicial equation has roots $s_1=1$ and $s_2=-1$. For $s_1=1$, the recurrence relation is:
\begin{equation}
    a_n=\dfrac{A_2}{A_0}\dfrac{a_{n-2}}{n+2}\qquad \text{for}\qquad n\geq2,
\end{equation}
which gives the first Frobenius solution:
\begin{equation}
    r_{s_1}(R)=R\,\sum_{k=0}^{\infty}\left(\dfrac{A_2}{2A_0}\right)^k\dfrac{R^{2k}}{(k+1)!}=\dfrac{2A_0}{A_2}\dfrac{(\mathrm{e}^{A_2R^2 \slash 2A_0}-1)}{R}.
\end{equation}
Since the two roots are separated by an integer, the second Frobenius solution can be obtained by using the reduction of order formula~\cite{Kreyszig11}:
\begin{equation}
    r_{s_2}(R)= r_{s_1}(R)\int_{}^{R}\dfrac{1}{r_{s_1}^2(u)}\mathrm{e}^{-\displaystyle\int^{u}\left[(A_0-A_2x^2) \slash A_0x\right]\dd{x}}\dd{u}=-\dfrac{1}{2R}.
\end{equation}
The homogeneous solution of~$\eqref{eq:gov0}_1$ is thus:
\begin{equation}
    r_1^{\mathrm{h}}(R)=d_1 r_{s_1}(R)+d_2 r_{s_2}(R)=d_1 \dfrac{2A_0}{A_2}\dfrac{(\mathrm{e}^{A_2R^2 \slash 2A_0}-1)}{R}-d_2\dfrac{1}{2R},
\end{equation}
and the particular solution can be easily found since the inhomogeneous term of~$\eqref{eq:gov0}_1$ is a polynomial of degree three in $R$ as follows:
\begin{equation}
    r_1^{\mathrm{p}}(R)=-\dfrac{(4 A_0 A_1 + 2 A_2 A_3) }{4 A_2^{2}}R- \frac{A_1}{4 A_2} R^3.
\end{equation}
The general solution is thus:
\begin{equation}
    r_1(R)=r_1^{\mathrm{h}}(R)+r_1^{\mathrm{p}}(R)=d_1 \dfrac{2A_0}{A_2}\dfrac{(\mathrm{e}^{A_2R^2 \slash 2A_0}-1)}{R}-d_2\dfrac{1}{2R}-\dfrac{(4 A_0 A_1 + 2 A_2 A_3)}{4 A_2^{2}}R- \frac{A_1}{4 A_2} R^3.\label{eq:genr1}
\end{equation}
Now, imposing the regularity condition $r_1(0)=0$ enforces $d_2=0$, while the boundary condition in~$\eqref{eq:gov0}_2$ gives:
\begin{equation}
    d_1=\frac{
\mathrm{e}^{-A_2R^2 \slash 2A_0}
\left[
2 A_0 A_3 + A_2A_1+(C_0A_2 + A_3)A_2 R_0^{2}
\right]
}{2 A_2^{2}}.\label{eq:d1}
\end{equation}
At order one, for compactness, the governing problem can be written in terms of $J_2$ as follows:
\begin{equation}
	\left\{
	\begin{aligned}
	& J_2^{\prime}(R)- \dfrac{A_2}{A_0} RJ_2(R)= P(R), \\
    & J_2(R_0)=C_1,
	\end{aligned}
    \right.
    \label{eq:gov1}
\end{equation}
where the constant $C_1$ is given in \eqref{app:C1} and the inhomogeneous term is given by:
\begin{equation}
    P(R)=\dfrac{p_0(R)
+ p_1(R) r_1(R)
+ p_2(R) r_1^{\prime}(R)
}{A_0}- \frac{5 A_2}{2 A_0}
RJ_1(R)^2.\label{eq:inhom1storder}
\end{equation}
The coefficients $p_i(R)$ are polynomial functions of $R$ and are listed in \eqref{app:p0}--\eqref{app:p2}. An analytical solution of~\eqref{eq:gov1} can be found by using the integrating factor method. Multiplying~$\eqref{eq:gov1}_1$ by the integrating factor $\mathrm{e}^{-\int^{R} \left(A_2u \slash A_0\right)\dd{u}}=\mathrm{e}^{-A_2R^2 \slash 2A_0}$, we obtain:
\begin{equation}
    \diff{}{R}\left(J_2(R)\mathrm{e}^{-A_2R^2 \slash 2A_0}\right)=\mathrm{e}^{-A_2R^2 \slash 2A_0}P(R).\label{interm}
\end{equation}
Integrating this equation and imposing the boundary condition~$\eqref{eq:gov1}_2$ yields the following solution:
\begin{equation}
    J_2(R)=-\mathrm{e}^{A_2R^2 \slash 2A_0}\int_{R}^{R_0} \mathrm{e}^{-A_2u^2 \slash 2A_0}P(u)\dd{u}+C_1\mathrm{e}^{A_2\left(R^2-R^2_0\right) \slash 2A_0}.\label{eq:J2sol}
\end{equation}
Although the integral in~\eqref{eq:J2sol} can be evaluated explicitly, the resulting expression is unwieldy and is therefore omitted for brevity. Finally, the expressions for $r_1(R)$ and $J_2(R)$ in~\eqref{eq:genr1}, \eqref{eq:d1} and~\eqref{eq:J2sol} can be substituted into~\eqref{eq:Tijform} to get the stress components $T_{\theta z}$ and $T_{zz}$. The resultant torque and axial force in~\eqref{eq:torqueforce} can be obtained, upon integration. Practically, we expand the solution $r_1(R)$ in~\eqref{eq:genr1}--\eqref{eq:d1} about $R=0$ up to order seven and similarly for $J_1(R)$ up to order six. We then use these expansions to calculate $P(R)$ in~\eqref{eq:inhom1storder} and substitute into~\eqref{eq:J2sol} to calculate the solution $J_2(R)$. Once both $r_1(R)$ and $J_2(R)$ are converted into integrable functions, we can then calculate the stress components $T_{\theta z}$ and $T_{zz}$ and finally evaluate the torque and axial force using~\eqref{eq:torqueforce}. 

%%%%%%%%%%%%%%%%%%%%%%%%%%%%%%%%%%%%%%%%%%%%%%%%%%%%%%
\section{Results and discussion}
\label{sec: Results and discussion}

The analytical expressions derived in Sections~\ref{sec: simple shear} and~\ref{sec:theotorsion} for the shear and normal stresses in simple shear and for the torque and axial force in torsion are valid for arbitrary shear and twist histories $k(t)$ and $\phi(t)$, respectively. We now consider a typical experimental ramp-and-hold loading scenario, for which the shear and twist histories are given by:
\begin{equation}
    k(t)=\frac{k_0}{t^{\star}}t-\frac{k_0}{t^{\star}}(t-t^{\star}) H(t-t^{\star}) \qquad \text{and} \qquad \phi(t)=\frac{\phi_0}{t^{\star}}t-\frac{\phi_0}{t^{\star}}(t-t^{\star}) H(t-t^{\star}), 
    \label{eq:input}
\end{equation} 
where $k_0$ and $\phi_0$ are, respectively, the maximum values of the shear and twist, $t^{\star}$ is the rising time of the ramp and $H$ is the Heaviside step function. Since ramp-phase data is often neglected in model fitting~\cite{matjeka26,small25}, we restrict our attention in this section to the hold phase (i.e.~$t \geq t^{\star}$) and plot the corresponding relaxation curves in simple shear and torsion to compare the predictions of the MQLV models developed previously in Sections~\ref{sec: simple shear} and~\ref{sec:theotorsion}. 

%%%%%%%%%%%%%%%%%%%%%%%%%%%%%%%%%%%%%%%%%%%%%%%%%%%%%%

\subsection{Effect of compressibility}
\label{sec: effect of compressibility}

\begin{figure}[t!]
    \centering
    \begin{subfigure}{0.49\textwidth}
        \centering
        \includegraphics[width=\linewidth]{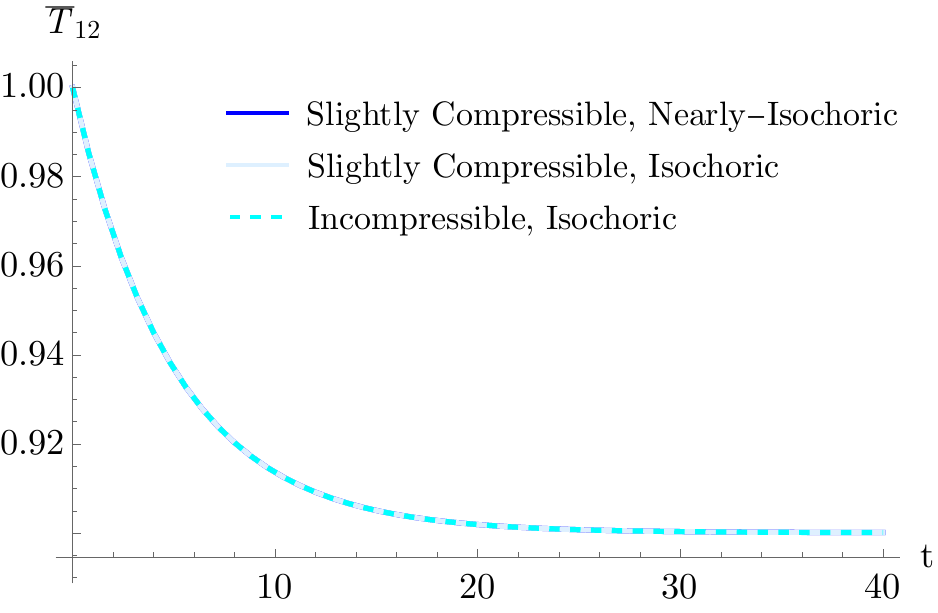}
        \caption{}
        \label{fig: compressibility shear stress}
    \end{subfigure}
    \hfill
    \begin{subfigure}{0.49\textwidth}
        \centering
        \includegraphics[width=\linewidth]{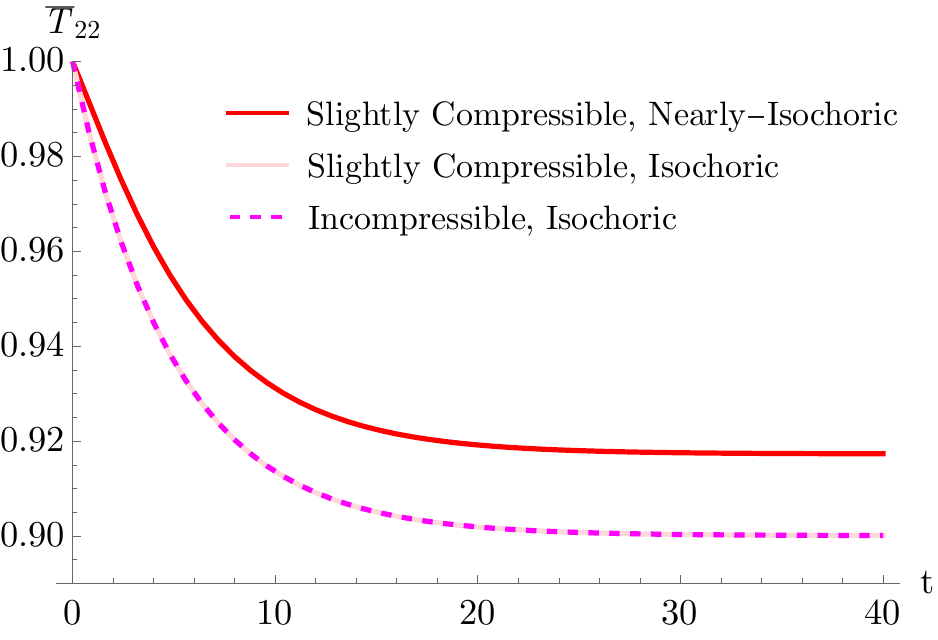}
        \caption{}
        \label{fig: compressibility normal stress}
    \end{subfigure}
    \begin{subfigure}{0.49\textwidth}
        \centering
        \includegraphics[width=\linewidth]{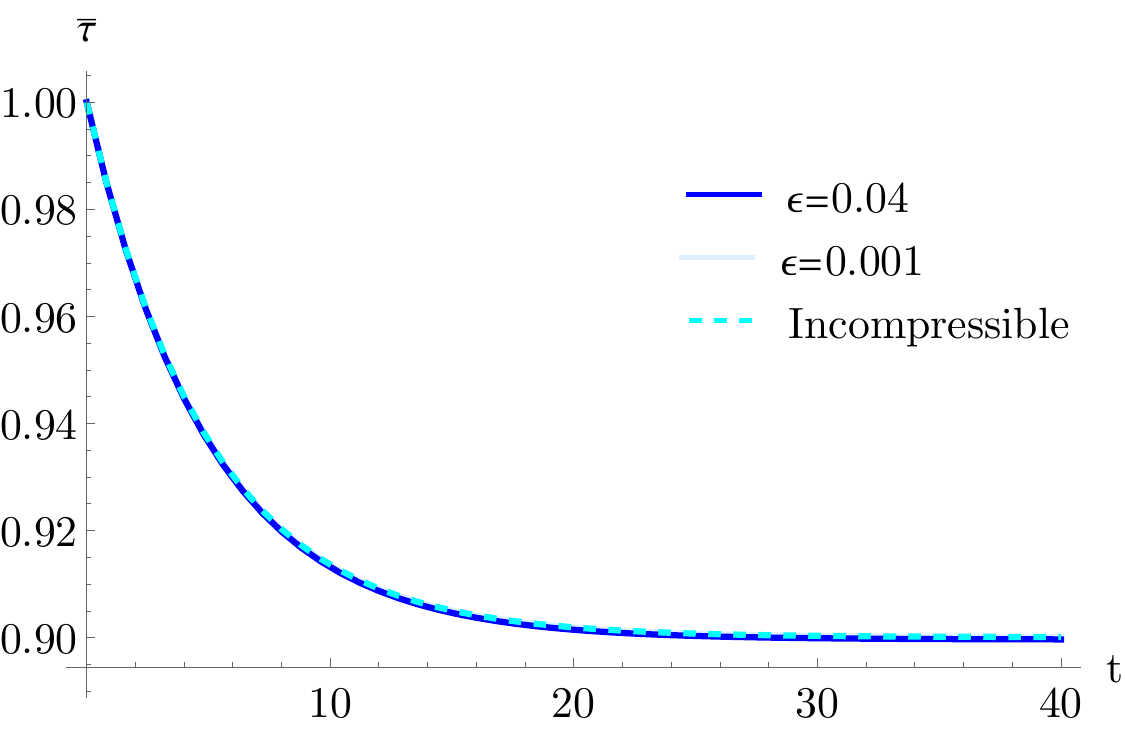}
        \caption{}
        \label{fig: compressibility torque}
    \end{subfigure}
    \begin{subfigure}{0.49\textwidth}
        \centering
        \includegraphics[width=\linewidth]{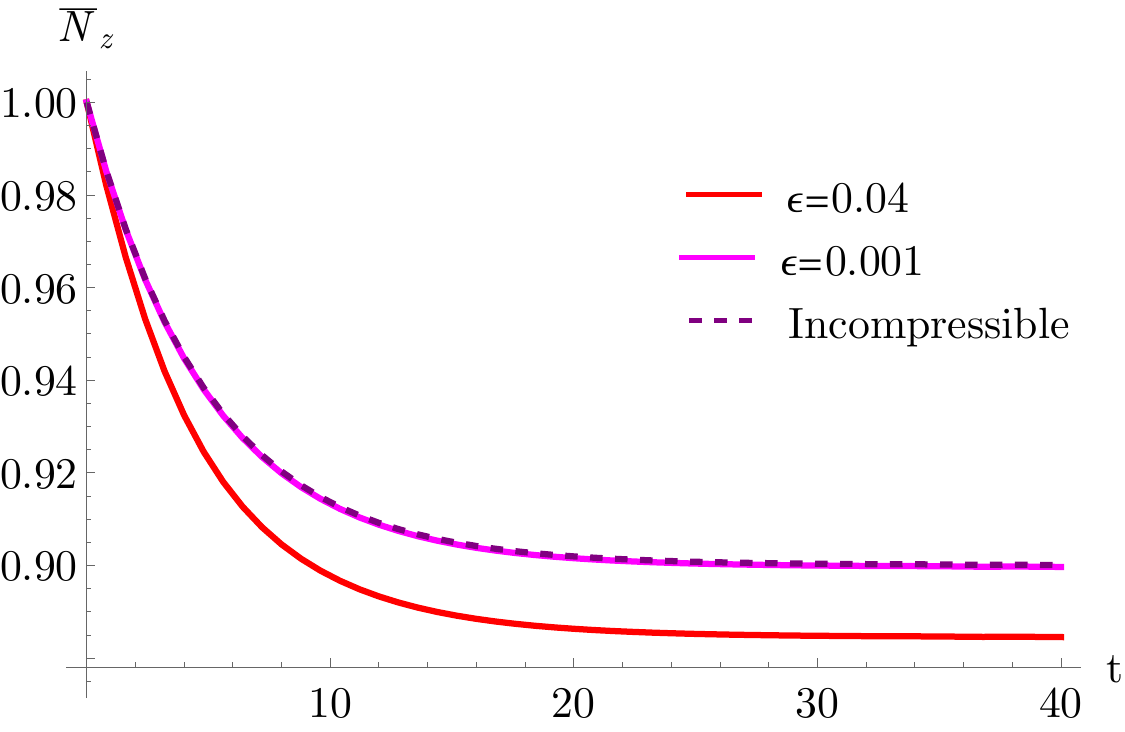}
        \caption{}
        \label{fig: compressibility force}
    \end{subfigure}
    \caption{Effect of compressibility: normalised shear stress $\overline{T}_{12}=T_{12}(t)/T_{12}(t^{\star})$ (a) and normal stress $\overline{T}_{22}=T_{22}(t)/T_{22}(t^{\star})$ (b) in simple shear for the isochoric deformation \eqref{Isochoric simple shear} (incompressible and slightly compressible constitutive models) and for the nearly isochoric deformation \eqref{Nearly isochoric simple shear} 
     (slightly compressible model); normalised torque $\overline{\tau}(t)/\overline{\tau}(t^\star)$ (c) and axial force $\overline{N}_z(t)/\overline{N}_z(t^\star)$ (c) in torsion. The following parameters are fixed: $k_0=0.4$, $\phi_0=25\,\mathrm{rad}\,\mathrm{m}^{-1}$, $R_0=12.5\,\mathrm{mm}$, $c_1=3000\,\mathrm{Pa}$, $c_2=2000\,\mathrm{Pa}$, $\mu_0=10^4\,\mathrm{Pa}$, $\tau_{\mathrm{D}}=5\,\mathrm{s}$, $\tau_{\mathrm{H}}=2\,\mathrm{s}$, $\kappa_{\infty}\slash \kappa_0=0.4$ and $\mu_{\infty}\slash\mu_0=0.9$. All curves in (a) and (b) are plotted for $\kappa_0=25\,\mu_0$. For the nearly-isochoric deformation in (a) and (b), we set $\delta=10^{-4}$. In (c) and (d), we vary $\varepsilon=\mu_0\slash\kappa_0 \in \{0.04,0.001\}$ and the incompressible curves are plotted from~\eqref{eq:tau} and~\eqref{eq:Fz}.}
    \label{fig: effect of compressibility}
\end{figure}

We begin by examining the effect of compressibility in the two deformation modes. Figures \ref{fig: compressibility shear stress} and \ref{fig: compressibility normal stress} show the relaxation of the shear and normal stresses for three simple shear scenarios: the incompressible and slightly compressible models under the isochoric deformation~\eqref{Isochoric simple shear}, and the slightly compressible model under the nearly isochoric deformation~\eqref{Nearly isochoric simple shear}. The corresponding torsional responses are displayed in Figures \ref{fig: compressibility torque} and \ref{fig: compressibility force} for different values of $\varepsilon=\mu_0/\kappa_0$.
In simple shear, the shear stress responses are identical in all three cases (Figure~\ref{fig: compressibility shear stress}), as can be verified by comparing \eqref{T12 isochoric simple shear incompressible final} and \eqref{T12 isochoric simple shear slightly compressible}. This follows from the isochoric nature of deformation~\eqref{Isochoric simple shear}: since $J=1$, the volumetric contribution vanishes and the pull-back of the volumetric stress $\boldsymbol{\Pi}^{\text{e}}_{\text{H}}$ in \eqref{Second Piola-Kirchhoff stress tensors slightly compressible} is zero.
In contrast, the normal stress exhibits a markedly different behaviour (Figure~\ref{fig: compressibility normal stress}). For the isochoric case, the incompressible and slightly compressible predictions \eqref{T22 isochoric simple shear incompressible final} and \eqref{T22 isochoric simple shear slightly compressible} differ only by a constant factor, so their normalised responses coincide (overlapping dashed and light-red curves). However, for the nearly isochoric deformation~\eqref{Nearly isochoric simple shear}, even a very small volumetric change ($\delta=10^{-4}$) leads to a substantial deviation in the normal stress relaxation (red curve), demonstrating the strong sensitivity of normal stresses to volumetric effects.

In torsion, the influence of compressibility is governed by $\varepsilon$. For $\varepsilon=0.001$, the slightly compressible response coincides with the incompressible one: both the torque (dashed and light-blue curves in Figure~\ref{fig: compressibility torque}) and the axial force (dashed and magenta curves in Figure~\ref{fig: compressibility force}) match the incompressible predictions given by~\eqref{eq:tau} and~\eqref{eq:Fz}. As $\varepsilon$ increases to $0.04$, the torque remains unaffected by compressibility. By contrast, the axial force becomes highly sensitive: its relaxation curve departs significantly from the incompressible prediction, highlighting the key role of volumetric effects in the axial response.

\subsection{Strain-dependence}
\begin{figure}[t!]
    \centering
    \begin{subfigure}{0.49\textwidth}
        \centering
        \includegraphics[width=\linewidth]{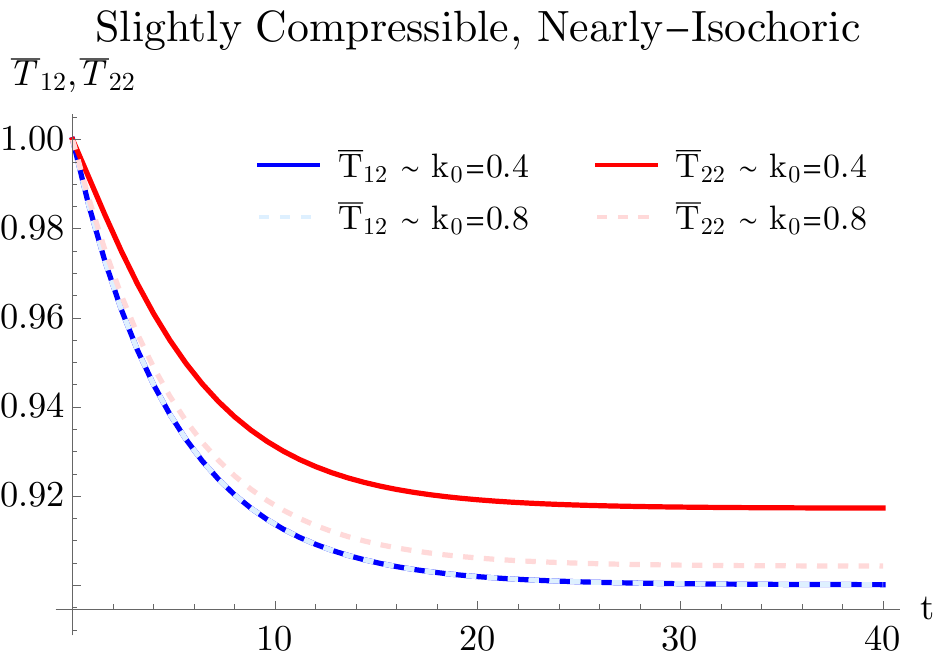}
        \caption{}
        \label{fig: effect of deformation isochoric shear}
    \end{subfigure}
    \hfill
    \begin{subfigure}{0.49\textwidth}
        \centering
        \includegraphics[width=\linewidth]{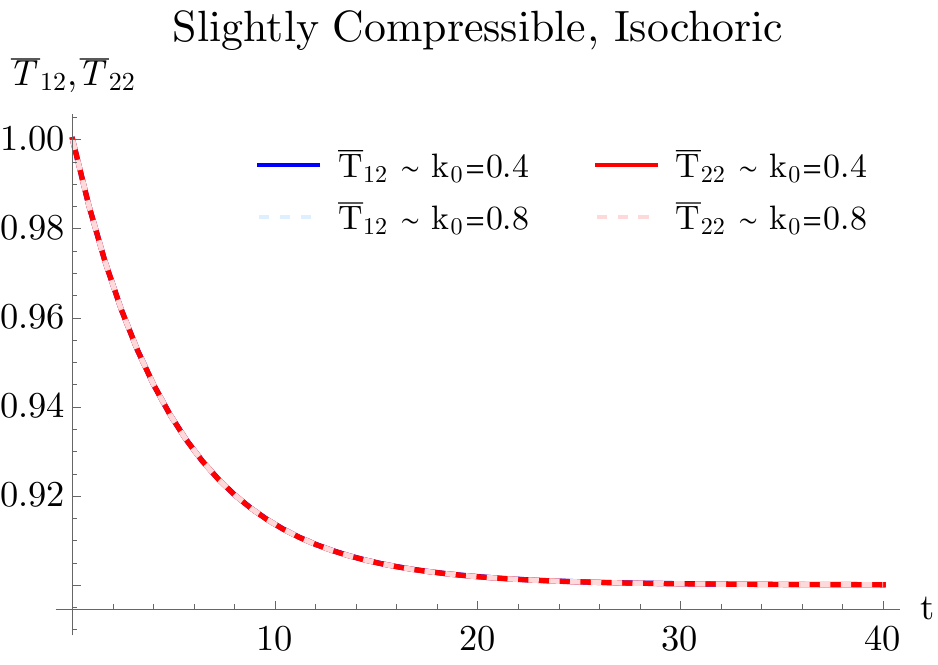}
        \caption{}
        \label{fig: effect of deformation nearly-isochoric shear}
    \end{subfigure}
    \begin{subfigure}{0.49\textwidth}
        \centering
        \includegraphics[width=\linewidth]{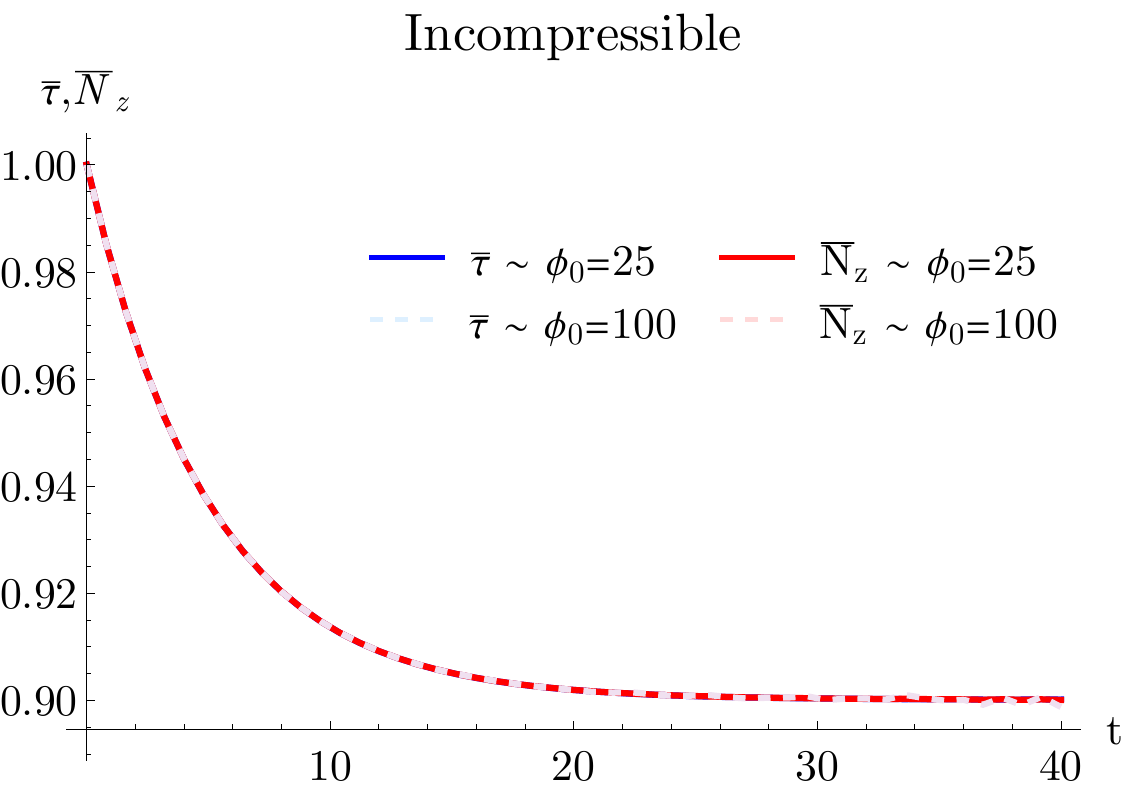}
        \caption{}
        \label{fig: effect of deformation incompressible torsion}
    \end{subfigure}
    \hfill
    \begin{subfigure}{0.49\textwidth}
        \centering
        \includegraphics[width=\linewidth]{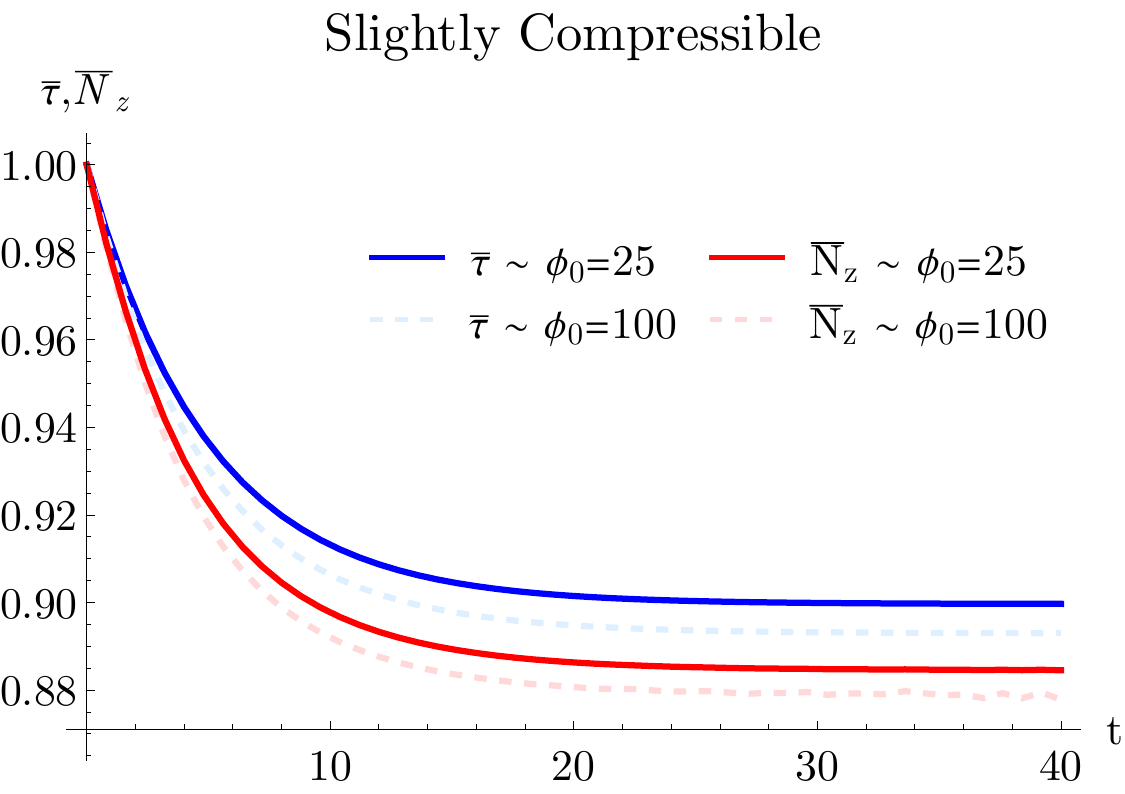}
        \caption{}
        \label{fig: effect of deformation slightly compressible torsion}
    \end{subfigure}
    \caption{Strain-dependence: normalised shear stress $\overline{T}_{12}$ and normal stress $\overline{T}_{22}$ predicted by the slightly compressible model (a) for the isochoric deformation \eqref{Isochoric simple shear} and (b) for the nearly isochoric deformation \eqref{Nearly isochoric simple shear}; normalised torque $\overline{\tau}$ and axial force $\overline{N}_{z}$ predicted by the incompressible model (c) and by the slightly compressible model (d) in torsion. The following parameters are fixed: $R_0=12.5\,\mathrm{mm}$, $c_1=3000\,\mathrm{Pa}$, $c_2=2000\,\mathrm{Pa}$, $\mu_0=10^4\,\mathrm{Pa}$, $\kappa_0=25\,\mu_0$, $\tau_{\mathrm{D}}=5\,\mathrm{s}$, $\tau_{\mathrm{H}}=2\,\mathrm{s}$, $\kappa_{\infty}\slash \kappa_0=0.4$, $\mu_{\infty}\slash\mu_0=0.9$ and $\delta=10^{-4}$. In (a) and (b), we vary $k_0 \in \{0.4,0.8\}$, while in (c) and (d) we vary $\phi_0 \in \{25,100\}\,\mathrm{rad}\,\mathrm{m}^{-1}$.}
    \label{fig: effect of large deformation}
\end{figure}

Next, we investigate whether compressibility amplifies the non-linear effects due to the large deformation, both in simple shear and in torsion. In light of its pronounced sensitivity to volume changes, we restrict attention to the slightly compressible model in simple shear. In Figure~\ref{fig: effect of large deformation}, we compare the normalised shear and normal stresses for the slightly compressible model under (a)~the isochoric simple shear deformation and (b)~the nearly-isochoric deformation at different levels of shear $k_0$. 
The curves show that, under the isochoric deformation, neither the shear stress nor the normal stress is affected by the magnitude of the applied shear. In contrast, for the non-isochoric deformation, the normal stress becomes sensitive to the level of shear. This behaviour arises because the bulk relaxation function is only activated when a volume change occurs. As a result, in the non-isochoric case, the bulk and shear contributions couple differently with the deformation, introducing an additional dependence of the normal stress on the applied shear.

In torsion, we compare the torque and axial force predicted by the incompressible (Figure~\ref{fig: effect of deformation incompressible torsion}) and slightly compressible (Figure~\ref{fig: effect of deformation slightly compressible torsion}) models for different levels of twist. We choose two levels of twist: $\phi_0=25\,\mathrm{rad}\,\mathrm{m}^{-1}$, which corresponds to a moderate twist, and $\phi_0=100\,\mathrm{rad}\,\mathrm{m}^{-1}$, which corresponds to a large twist. In the incompressible case, the normalised torque and axial force curves coincide, as both depend on a single relaxation function $\mu(t)$. By contrast, in the slightly compressible model, both the torque and axial force relaxation profiles change with the level of twist, thus exhibiting strain-dependent relaxation.

The strain-dependence observed in torsion and in the normal stress response in simple shear primarily stems from the different way in which the shear and bulk relaxation functions couple with the deformation. 
In simple shear, the hereditary integrals associated with $\mu(t)$ and $\kappa(t)$ in the shear stress combine in a factorisable manner with respect to $k(t)$, leading to uniform scaling that is removed upon normalisation. In contrast, the normal stress involves a non-factorisable combination of terms with different powers of $k(t)$, including contributions independent of $k(t)$, so that its normalised response retains a dependence on the deformation (compare \eqref{T12 nearly isochoric simple shear} and \eqref{T22 nearly isochoric simple shear}). Similarly, in torsion the coupling between the twist $\phi(t)$ and the relaxation functions $\mu(t)$ and $\kappa(t)$ is non-factorisable, so that the relative contributions of shear and bulk relaxation depend on the level of twist, leading to strain-dependent behaviour in both torque and axial force.
\subsection{Competition between shear and bulk relaxation}
\begin{figure}[h!]
    \centering
    \begin{subfigure}{0.49\textwidth}
        \centering
        \includegraphics[width=\linewidth]{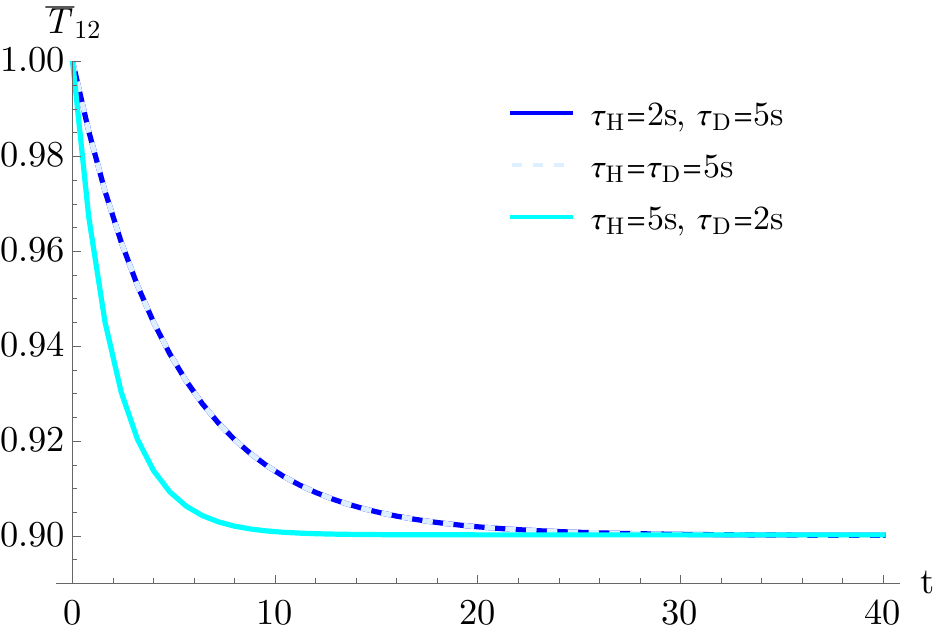}
        \caption{}
        \label{fig: effect of tau shear stress}
    \end{subfigure}
    \hfill
    \begin{subfigure}{0.49\textwidth}
        \centering
        \includegraphics[width=\linewidth]{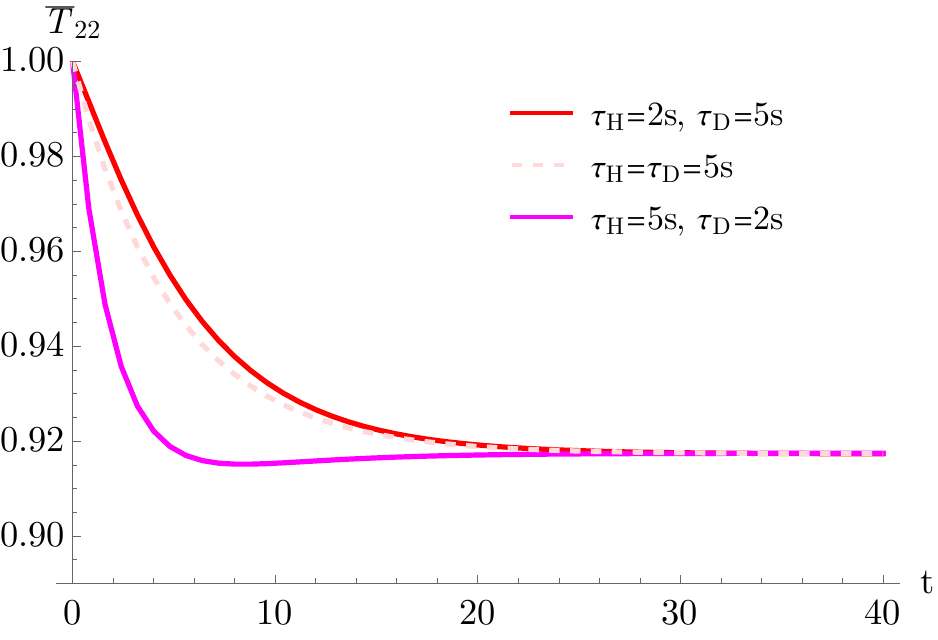}
        \caption{}
        \label{fig: effect of tau normal stress}
    \end{subfigure}
    \begin{subfigure}{0.49\textwidth}
        \centering
        \includegraphics[width=\linewidth]{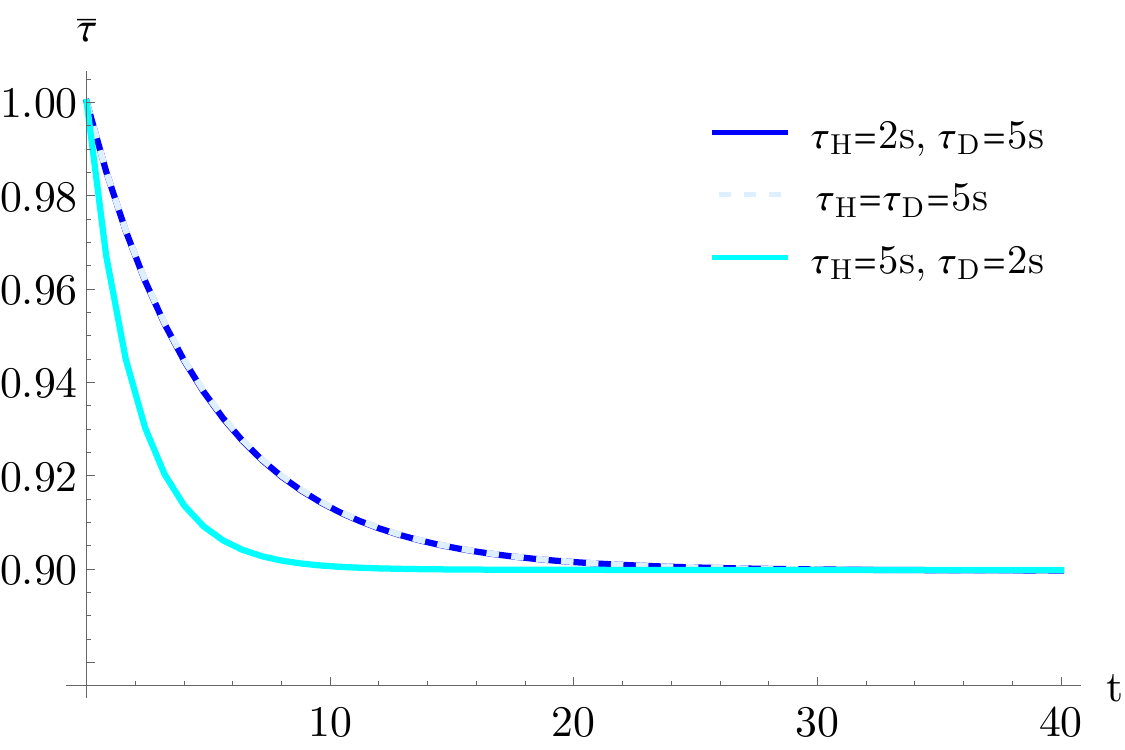}
        \caption{}
        \label{fig: effect of tau torque}
    \end{subfigure}
    \hfill
    \begin{subfigure}{0.49\textwidth}
        \centering
        \includegraphics[width=\linewidth]{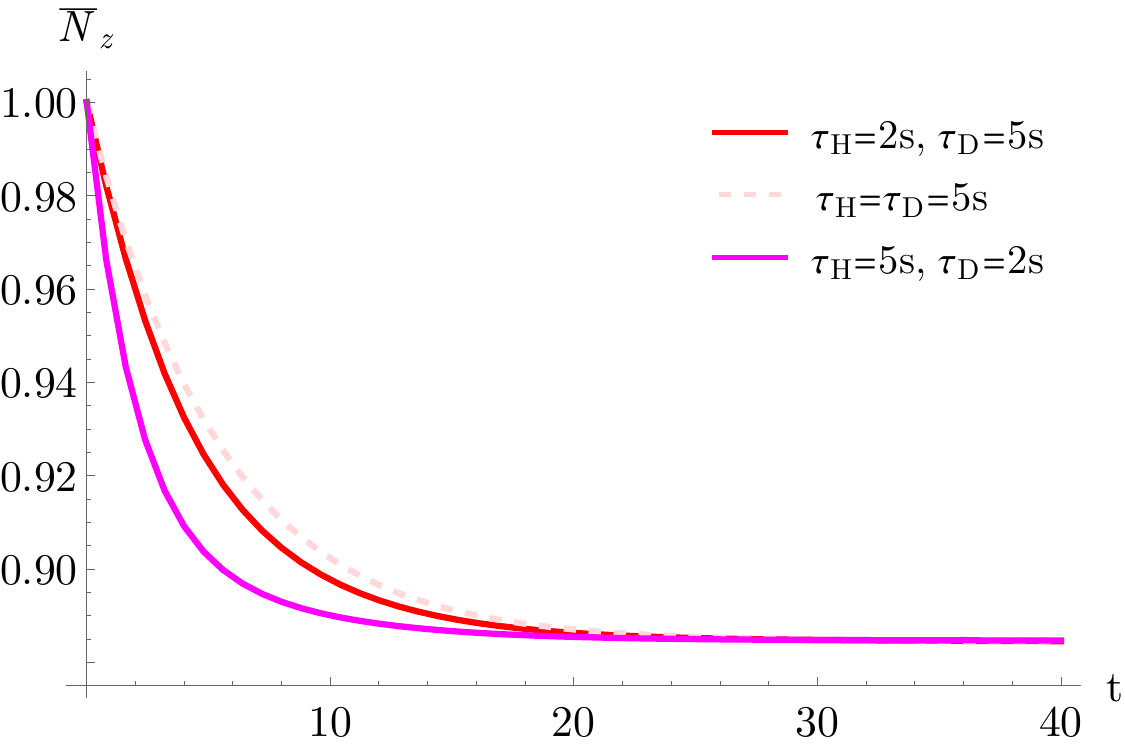}
        \caption{}
        \label{fig: effect of tau force}
    \end{subfigure}
    \caption{Effect of the relaxation times $\tau_{\mathrm{D}}$ and $\tau_{\mathrm{H}}$: normalised shear stress $\overline{T}_{12}$ (a) and normal stress $\overline{T}_{22}$ (b) predicted by the slightly compressible model for the nearly-isochoric deformation \eqref{Nearly isochoric simple shear}; normalised torque $\overline{\tau}$ (c) and axial force $\overline{N}_z$ (d) predicted by the slightly compressible model. The following parameters are fixed: $R_0=12.5\,\mathrm{mm}$, $c_1=3000\,\mathrm{Pa}$, $c_2=2000\,\mathrm{Pa}$, $\mu_0=10^4\,\mathrm{Pa}$, $\kappa_0=25\,\mu_0$, $\kappa_{\infty}\slash \kappa_0=0.4$, $\mu_{\infty}\slash\mu_0=0.9$, $\delta=10^{-4}$ and $k_0=0.4$. We vary the following pair of parameters: $\tau_{\mathrm{H}}=2\,\mathrm{s}$, $\tau_{\mathrm{D}}=5\,\mathrm{s}$ (blue and red curves), $\tau_{\mathrm{H}}=\tau_{\mathrm{D}}=5\,\mathrm{s}$ (dashed curves), $\tau_{\mathrm{H}}=5\,\mathrm{s}$, $\tau_{\mathrm{D}}=2\,\mathrm{s}$ (cyan and magenta curves).}
    \label{fig: effect of relaxation times}
\end{figure}
\begin{figure}[b!]
    \centering
    \begin{subfigure}{0.49\textwidth}
        \centering
        \includegraphics[width=\linewidth]{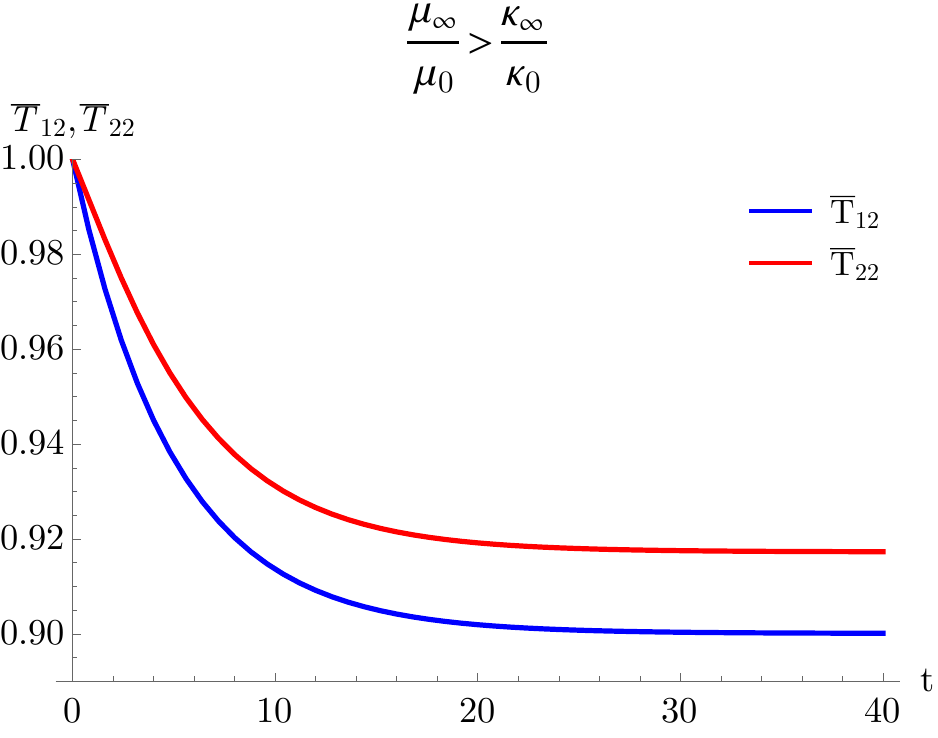}
        \caption{}
        \label{fig: simple shear MUinf greater}
    \end{subfigure}
    \hfill
    \begin{subfigure}{0.49\textwidth}
        \centering
        \includegraphics[width=\linewidth]{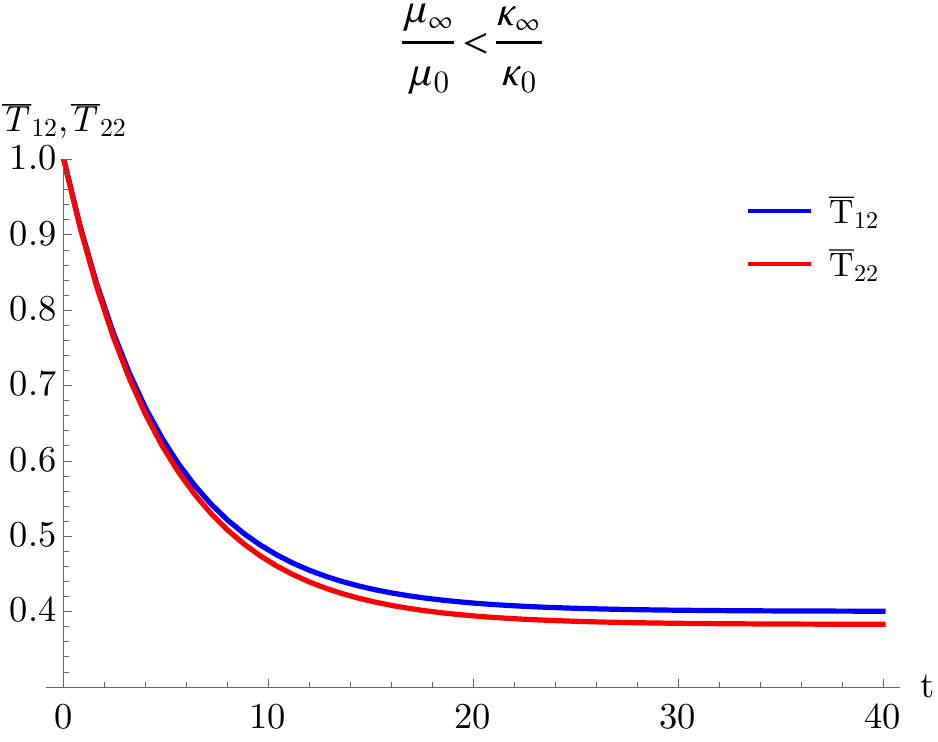}
        \caption{}
        \label{fig: simple shear MUinf smaller}
    \end{subfigure}
    \begin{subfigure}{0.49\textwidth}
        \centering
        \includegraphics[width=\linewidth]{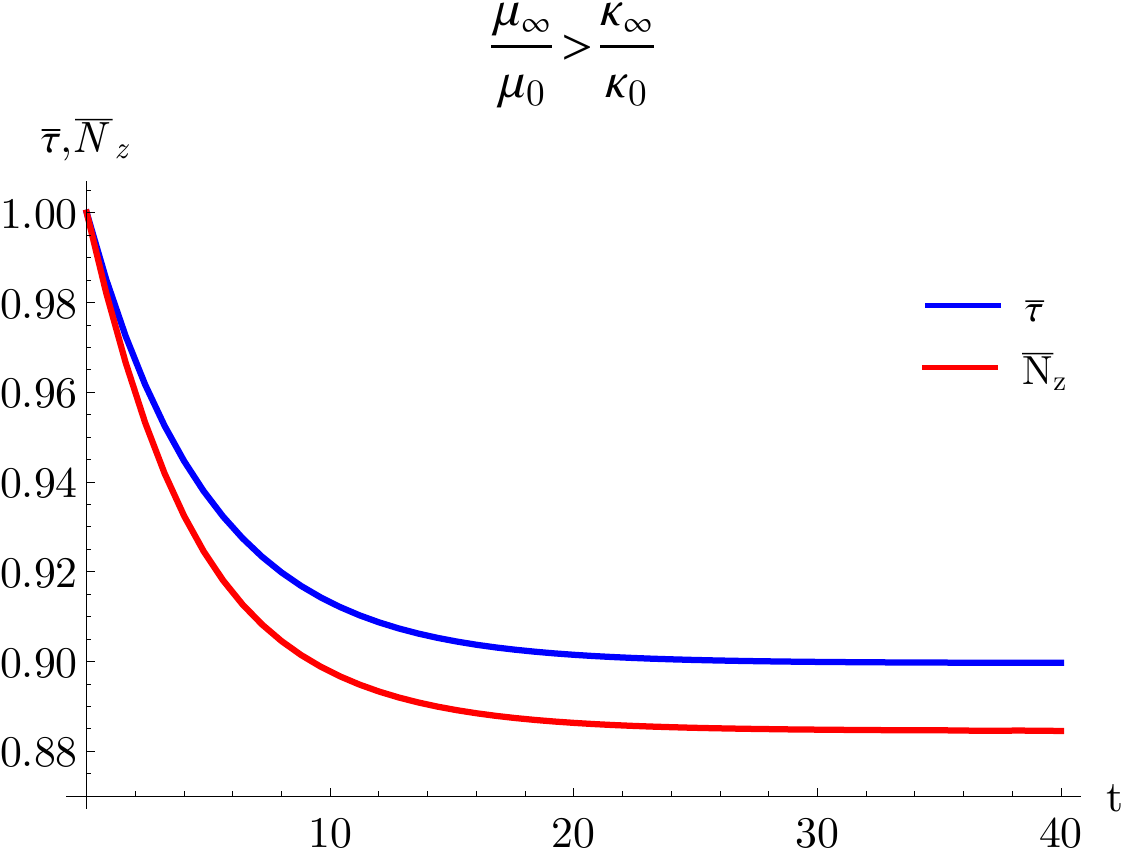}
        \caption{}
        \label{fig: torsion MUinf greater}
    \end{subfigure}
    \hfill
    \begin{subfigure}{0.49\textwidth}
        \centering
        \includegraphics[width=\linewidth]{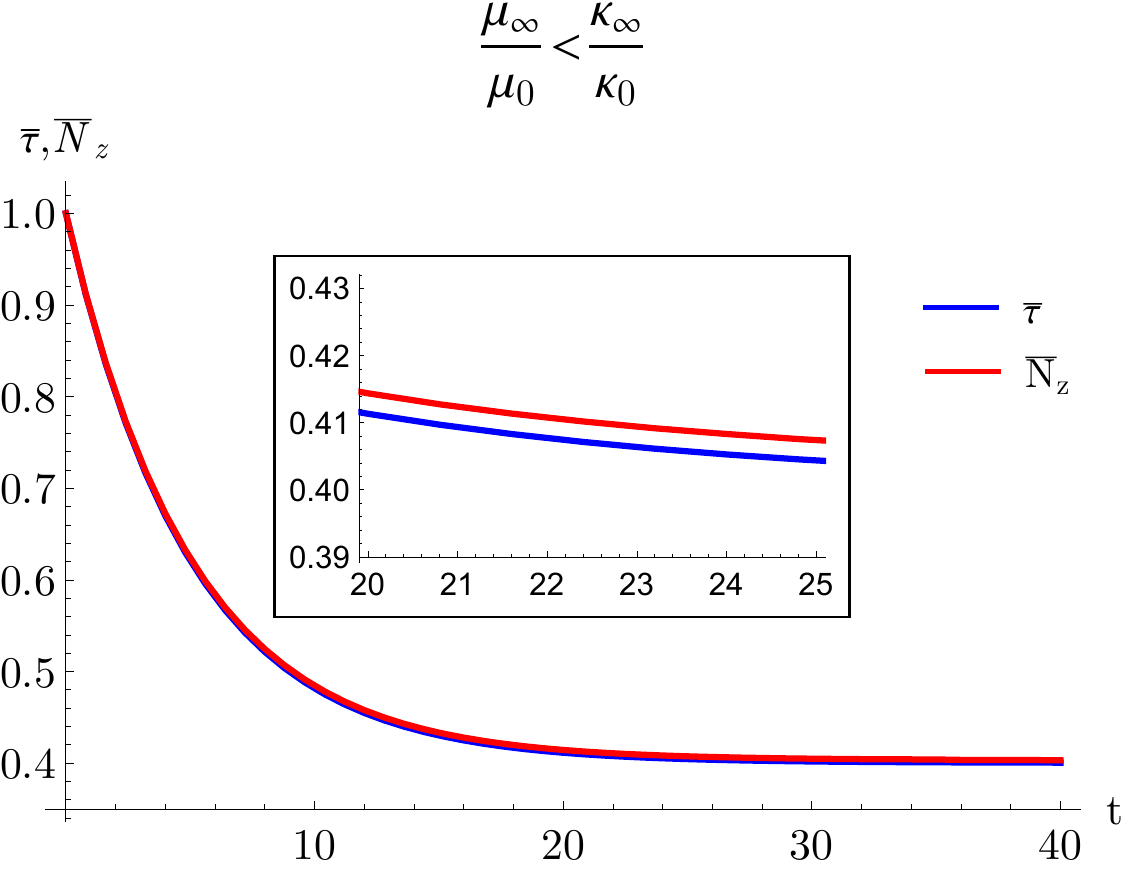}
        \caption{}
        \label{fig: torsion MUinf smaller}
    \end{subfigure}
    \caption{Effect of normalised long-time moduli $\mu_\infty\slash\mu_0$ and $\kappa_\infty\slash\kappa_0$: (a) and (b) normalised shear and normal stresses predicted by the slightly compressible model for the nearly-isochoric deformation \eqref{Nearly isochoric simple shear} in simple shear; (c) and (d) normalised torque and axial force in torsion predicted by the slightly-compressible model. The following parameters are fixed: $R_0=12.5\,\mathrm{mm}$, $c_1=3000\,\mathrm{Pa}$, $c_2=2000\,\mathrm{Pa}$, $\mu_0=10^4\,\mathrm{Pa}$, $\kappa_0=25\,\mu_0$, $\tau_{\mathrm{D}}=5$s, $\tau_{\mathrm{H}}=2$s, $\delta=10^{-4}$ and $k_0=0.4$. In (a) and (c) we set: $\mu_\infty\slash\mu_0=0.9$, $\kappa_\infty\slash\kappa_0=0.4$, while in (b) and (d) we set: $\mu_\infty\slash\mu_0=0.4$, $\kappa_\infty\slash\kappa_0=0.9$.}
    \label{fig: effect of normalised moduli}
\end{figure}

Next, in Figures~\ref{fig: effect of relaxation times} and \ref{fig: effect of normalised moduli}, we quantify the interplay between the shear and bulk relaxation functions $\mu(t)$ and $\kappa(t)$ by varying the relaxation times $\tau_{\mathrm{D}}$, $\tau_{\mathrm{H}}$ and the normalised long-term moduli $\mu_\infty/\mu_0$, $\kappa_\infty/\kappa_0$.
In Figure~\ref{fig: effect of relaxation times}, we consider $\tau_{\mathrm{D}} > \tau_{\mathrm{H}}$, $\tau_{\mathrm{D}} = \tau_{\mathrm{H}}$, and $\tau_{\mathrm{D}} < \tau_{\mathrm{H}}$.

The relaxation times primarily affect the short-time behaviour. In simple shear, the shear stress is insensitive to changes in $\tau_{\mathrm{H}}$, as confirmed by the overlap of curves with identical $\tau_{\mathrm{D}}$ in Figure~\ref{fig: effect of tau shear stress}. The same behaviour is observed for the torque in Figure \ref{fig: effect of tau torque}. In contrast, both normal stress and axial force (Figures \ref{fig: effect of tau normal stress} and \ref{fig: effect of tau force}) noticeably vary with $\tau_{\mathrm{H}}$, highlighting a competition between shear and bulk relaxation mechanisms. 
Figures~\ref{fig: simple shear MUinf greater} and~\ref{fig: simple shear MUinf smaller} again confirm that the dominant contribution to the shear stress arises from the shear relaxation function, since varying the ratio $\mu_{\infty}/\mu_0$ directly affects the asymptotic value of the relaxation curve. The plots also show that when the bulk relaxation effects are comparatively stronger, the relaxation behaviour of the normal stress deviates from that of the shear stress (compare Figure~\ref{fig: simple shear MUinf greater} with~\ref{fig: simple shear MUinf smaller}). A similar behaviour is observed in torsion (Figures~\ref{fig: torsion MUinf greater} and~\ref{fig: torsion MUinf smaller}).

Moreover, when the normalised long-time modulus of the shear relaxation function is greater than the bulk one, the axial force relaxes more than the torque (see Figure~\ref{fig: torsion MUinf greater}). However, when $\mu_\infty/\mu_0< \kappa_\infty/\kappa_0$ (i.e.~the relaxation associated with the shear relaxation function is greater) the torque relaxes more than the force (see Figure~\ref{fig: torsion MUinf smaller}). This latter behaviour is consistent with the experimental data shown in Figure~\ref{fig: brain data} for torsion tests on brain tissue samples. While the slightly compressible theory that we proposed in this paper is able to capture the relaxation behaviour of brain tissue samples, the torque and axial force curves are visibly different for the agarose gel samples (see Figure~\ref{fig: gel data}). This discrepancy is possibly due to differences in the underlying material structure. Although brain tissue consists of approximately $80\%$ water by volume, only $20$--$40\%$ of this is free flowing, with the remainder trapped inside the cells that comprise the solid matrix~\cite{Budday20}. By contrast, agarose gels can contain upwards of $90\%$ free-flowing water by volume and form a more open porous network~\cite{wang21}. As a result, agarose gels can undergo larger volume changes during deformation than brain tissue, which are not fully captured by the slightly compressible model proposed in this paper. For agarose gels, a fully compressible model may therefore be more appropriate. However, developing such a model for torsion is not straightforward, as torsion is not a universal deformation. From a mathematical perspective, in contrast to the incompressible case, where a deformation with $r=R$ is admissible for a wide class of materials, compressible elasticity generally requires additional radial deformation in order to satisfy the equation of motion. As a result, such a deformation is not compatible with all strain energy functions~\cite{kirkinis02,polignone91} and pure torsion is only recovered for special material choices. In the slightly compressible setting, this difficulty can be addressed perturbatively by introducing a small parameter $\varepsilon = \mu_0 \slash \kappa_0$ and expanding the deformation and governing equations accordingly, leading to a tractable approximate solution obtained via a linearisation of the governing equations at successive orders~\cite{small24,levinson1972}. However, in a fully compressible model, no such simplification is available and the coupled non-linear boundary value problem must be solved in full, with the resulting deformation depending explicitly on both the constitutive law and the geometry.

Finally, by comparing Figures~\ref{fig: simple shear MUinf greater} and \ref{fig: torsion MUinf greater}, an opposite trend emerges: in simple shear the normal stress relaxes less than the shear stress, whereas in torsion the axial force relaxes more than the torque. This opposite behaviour can be understood by examining the interplay between volumetric effects and the Poynting effect~\cite{horgan25,destrade25,zurlo20,poynting1909}. In simple shear, a slightly compressible material undergoes a small volume increase (i.e. $\delta>0$) associated with an elongation in the shear direction and contraction in the normal direction. This induces a positive normal stress contribution arising from compressibility, which opposes the negative normal stress generated by the Poynting effect. As a result, these two mechanisms compete, reducing the overall magnitude of relaxation and leading to a higher long-time value of the normal stress compared to the shear stress. In torsion, by contrast, the deformation induces a slight volume decrease. The resulting volumetric contribution generates a negative axial force, which adds to the negative contribution associated with the Poynting effect. In this case, the two mechanisms act cooperatively, enhancing the overall relaxation and leading to a lower long-time value of the axial force compared to the torque.

%%%%%%%%%%%%%%%%%%%%%%%%%%%%%%%%%%%%%%%%%%%%%%%%%%%%%%

\section{Conclusions}
\label{sec: Conclusion}

In this work, we have investigated the role of compressibility within the MQLV framework.
Our results showed that compressibility is only activated through the bulk relaxation function $\kappa(t)$ in the presence of a volume change. The bulk contribution vanishes under isochoric deformations, so that slightly compressible and incompressible models yield identical responses. Even a very small deviation from volume preservation, however, is sufficient to trigger bulk relaxation and significantly alter the stress response.

Second, the sensitivity to compressibility depends strongly on the stress measure. The shear stress in simple shear and the torque in torsion are only marginally affected by compressibility, as they are dominated by the shear relaxation function $\mu(t)$. In contrast, the normal stress in simple shear and the axial force in torsion exhibit a pronounced dependence on compressibility, reflecting the contribution of the bulk relaxation mechanism even under nearly-isochoric deformations.

Third, the slightly compressible MQLV model reveals a clear interplay between volumetric effects and the Poynting effect. In simple shear, the deformation induces a small volume increase, generating a positive volumetric contribution that opposes the negative Poynting effect; these competing mechanisms reduce the overall relaxation and increase the long-term value of the normal stress. In torsion, by contrast, the deformation leads to a volume decrease, and the resulting volumetric contribution reinforces the Poynting effect. This additive behaviour enhances the relaxation of the axial force, leading to a lower long-term value compared to the torque.

Finally, these findings have direct implications for the interpretation of experimental data. In particular, normal stress measurements such as the axial force in torsion or the normal stress in shear are significantly more sensitive to compressibility than the shear stress or torque. As a result, neglecting compressibility may lead to misinterpretation of relaxation curves and inaccurate identification of material parameters, especially when small but non-negligible volume changes are present. This highlights the importance of incorporating compressibility effects when fitting constitutive models to experimental observations involving normal stress responses.

Future work will focus on implementing the MQLV framework within finite element formulations, enabling the analysis of more general geometries, material heterogeneity and fully compressible behaviours beyond the semi-analytical approach considered here.

%%%%%%%%%%%%%%%%%%%%%%%%%%%%%%%%%%%%%%%%%%%%%%%%%%%%%%
% \enlargethispage{20pt}

\section*{Data access}
The brain and agarose gel torsion data referenced in Figure~\ref{fig:Normalised experimental data} are available in the Mendeley Data repositories:~\href{https://doi.org/10.17632/m2jwdfgczs.1}{https://doi.org/10.17632/m2jwdfgczs.1} and
\href{https://doi.org/10.17632/m2jwdfgczs.1}{https://doi.org/10.17632/ny3k9s7494.1}, respectively.

\section*{AI declaration}
In preparing this article, the authors utilised the ChatGPT large language model as a copy editing tool to improve grammar, refine language and enhance the clarity of the original text.

\section*{Authors contribution}
Valentina Balbi: conceptualisation, methodology, formal analysis, writing original draft, writing reviewing and editing. Griffen Small: conceptualisation, methodology, formal analysis, writing original draft, writing reviewing and editing. Both authors gave final approval for publication.

\section*{Competing Interests}
The authors declare that they have no known competing financial interests or personal relationships that could have appeared to influence the work reported in this article.

\section*{Funding}
This article has emanated from research jointly funded by the College of Science and Engineering at the University of Galway under the Millennium Fund scheme for the project ``Modelling Brain Mechanics'' (Valentina Balbi) and Taighde \'Eireann – Research Ireland under grant number GOIPG/2024/3552 (Griffen Small).

% \ack{}

%%%%%%%%%%%%%%%%%%%%%%%%%%%%%%%%%%%%%%%%%%%%%%%%%%%%%%
\appendix
\renewcommand{\thesection}{Appendix~\Alph{section}.}
\renewcommand{\theequation}{\Alph{section}.\arabic{equation}}
\setcounter{equation}{0}

\section{Hereditary integral terms}\label{App:integrals}
The following compact notation for the integrals $\mathrm{DInt}_i(t)$ and $\mathrm{HInt}_i(t)$ is used throughout the manuscript:
\begin{equation}
    \mathrm{DInt}_i(t)=-\frac{1}{\mu_0}\int_{0}^{t}\mu^{\prime}(t-s)x(s)^i\dd{s} \qquad \text{for} \qquad i=0,1,2,3,4. 
    \label{Deviatoric integrals shortand notation}
\end{equation}
and
\begin{equation}
    \mathrm{HInt}_i(t)=-\frac{1}{\kappa_0}\int_{0}^{t}\kappa^{\prime}(t-s)x(s)^i\dd{s} \qquad \text{for} \qquad i=0,1,2. 
    \label{Hydrostatic integrals shortand notation}
\end{equation}
where $x(s)=k(s)$ in simple shear and $x(s)=\phi(s)$ in torsion, according to \eqref{eq:input}.

\section{Governing equations for the slightly compressible model of torsion}\label{App:A}
The coefficients $A_i(t)$ appearing in the zeroth- and first-order governing equations~\eqref{eq:gov0} and~\eqref{eq:gov1}, respectively, are given by the following expressions:
\begin{align}
    &A_0 =\mu_0(\mathrm{HInt}_0(t)-1),\label{app:A0}\\
    &A_2=\mu_0(\mathrm{HInt}_0(t) \phi(t)^2- 2 \mathrm{HInt}_1(t) \phi(t)+\mathrm{HInt}_2(t)).\label{app:A2}
\end{align}
The coefficients $A_1$ and $A_3$ of the polynomial inhomogeneous term in the zeroth-order equation are:
\begin{align}
    &A_1=- \dfrac{2}{3} (5 c_1 - 2 c_2)\phi(t)^2+2\mu_0\phi(t)\mathrm{DInt}_1(t)-\dfrac{2}{3} (c_1 + 8 c_2)\mathrm{DInt}_2(t),\label{app:A1}\\
    &A_3=-\dfrac{2}{3} (c_1 + 2 c_2)\Bigl(\mathrm{DInt}_2(t)\phi(t)^2-2\mathrm{DInt}_3(t)\phi(t)+ \mathrm{DInt}_4(t)\Bigr).\label{app:A3}
\end{align}
The coefficients $p_0$, $p_1$ and $p_2$ of the inhomogeneous term in the first-order equation are:
\begin{alignat}{2}
&p_0&&=-\frac{R(A_1 + A_3 R^{2})}{3 A_0 }
\left[
A_0 C_0 \dfrac{(5 c_1 - c_2)}{(c_1 - c_2)} R^{2}
+ 4 \mu_0(\mathrm{DInt}_0(t)-1)
\right], 
\label{app:p0} \\[5pt]
&p_1&&=
-\dfrac{R^{2}}{3}\left[
\frac{A_2 C_0 (5 c_1 - c_2)}{c_1 - c_2}
- \frac{A_3 (13 c_1 + 22 c_2)}{c_1 + 2 c_2}
\right] \nonumber \\
&&&+\dfrac{1}{3}\left[A_1
\left(
7 + \frac{2 c_1}{c_1 + c_2}
\right)
+ A_0 C_0
\left(
-4 + \frac{14 c_1 c_2}{c_1^{2} - c_2^{2}}
\right)-2\mu_0(\mathrm{DInt}_0(t)-1)
\Bigl(
2 \dfrac{A_2}{A_0} + \phi(t)^2
\Bigr)\right],
\label{app:p1} \\[-5pt]
&p_2&&=-\frac{R^{3}}{3}
\left[
\frac{A_2 C_0 (5 c_1 - c_2)}{c_1 - c_2}
- A_3 \left( 11 - \frac{4 c_1}{c_1 + 2 c_2} \right)
\right] \nonumber \\
&&&+\dfrac{R}{3}\left[\frac{A_1 (9 c_1 + 13 c_2)}{c_1 + c_2}
- 2 A_0 C_0
\left(
8 - \frac{2 c_1 (3 c_1 - 4 c_2)}
{c_1^2 - c_2^2}\right)-2\mu_0(\mathrm{DInt}_0(t)-1)
\Bigl(
2 \dfrac{A_2}{A_0} + \phi(t)^2
\Bigr)\right].
\label{app:p2}
\end{alignat}

The constants $C_0$ and $C_1$ in the boundary conditions at the zeroth- and first-orders are:
\begin{align}
    &C_0=\frac{2}{3}\frac{(c_1 - c_2)}{A_0}
\bigl(\mathrm{DInt}_2(t) - \phi(t)^2\bigr),\label{app:C0}\\
    &C_1=C_0R_0\left[\frac{2 c_1}{c_1 - c_2} r_1(R_0)-\frac{(c_1 + 7 c_2)}{6(c_1 - c_2)} C_0R_0^{3}\right]+\frac{2\mu_0(1 - \mathrm{DInt}_0(t))}{3 A_0R_0}
\Bigl(
2 C_0 R_0^{3} - 3 r_1(R_0)
\Bigr).\label{app:C1}
\end{align}

%%%%%%%%%% Insert bibliography here %%%%%%%%%%%%%%

% \vskip2pc

% \bibliographystyle{plain}
% \bibliography{Bibliography} % your .bib file name WITHOUT .bib

\end{document}